%% file: Main.tex
\documentclass[lettersize,journal]{IEEEtran}
\usepackage[utf8]{inputenc}
\usepackage{amsmath,amsfonts}
\usepackage{algorithmic}
\usepackage{array}
\usepackage[caption=false,font=normalsize,labelfont=sf,textfont=sf]{subfig}
\usepackage{textcomp}
\usepackage{stfloats}
\usepackage{url}
\usepackage{verbatim}
\usepackage{graphicx}
\usepackage{cite}
\usepackage[normalem]{ulem}

\def\BibTeX{{\rm B\kern-.05em{\sc i\kern-.025em b}\kern-.08em
    T\kern-.1667em\lower.7ex\hbox{E}\kern-.125emX}}
\usepackage{balance}

\usepackage[linesnumbered,ruled,vlined,english]{algorithm2e}
\usepackage{setspace}
\usepackage{lettrine}
\usepackage{xcolor}
\usepackage{arydshln}
\usepackage{multirow}
\usepackage{graphicx}
\usepackage{subfig}
\usepackage{booktabs}

\usepackage{orcidlink}

\begin{document}
\title{TetrisG-SDK: Efficient Convolutional Layer \\ Mapping with Adaptive Windows and Grouped\\ Convolutions for Fast In-Memory Computing}
\author{
    Ke Dong~\textsuperscript{\orcidlink{0000-0002-8973-9797}}, ~\IEEEmembership{Student Member,~IEEE}%
    \thanks{Ke Dong and Bo Wang are with the Information Systems Technology and Design, Singapore University of Technology and Design, Singapore 487372. E-mail: \{dong\_ke, bo\_wang\}@sutd.edu.sg. Kejie Huang is with the College of Information Science and Electronic Engineering, Zhejiang University, China 310027. E-mail: huangkejie@zju.edu.cn. Tao Luo is with A*STAR Institute of High Performance Computing, Singapore 138632. E-mail:luo\_tao@ihpc.a-star.edu.sg. This work was supported by MOE Singapore Academic Research Fund Tier 2 (Award No. MOE-T2EP50122-0024).}; 
    Kejie Huang~\textsuperscript{\orcidlink{0000-0003-3722-9979}},~\IEEEmembership{Senior Member,~IEEE}; %
    Tao Luo~\textsuperscript{\orcidlink{0000-0002-3415-3676}},~\IEEEmembership{Senior Member,~IEEE}; %
    Bo Wang~\textsuperscript{\orcidlink{0000-0001-9199-0799}},~\IEEEmembership{Senior Member,~IEEE}%
}


\markboth{}%
{Shell \MakeLowercase{\textit{et al.}}: TetrisG-SDK: Efficient Convolutional Layer Mapping with Adaptive Windows for Fast In-Memory Computing}



\maketitle

\begin{abstract}
Shifted‑and‑Duplicated‑Kernel (SDK) mapping has emerged as an effective strategy to accelerate convolutional layers on compute‑in‑memory (CIM) hardware.
However, existing SDK variants (e.g., VWC‑SDK) merely optimize mapping for a single CIM macro, leaving inter-macro parallelism unexplored.
Moreover, their mapping methodologies are still suboptimal.
To address these limitations, we present TetrisG-SDK, a novel framework that employs adaptive windows to boost mapping performance. The proposed windows accommodate more input channels, increase array utilization at marginal space, and adapt to different channel depths. 
More importantly, TetrisG-SDK reduces compute latency by searching for optimal window configurations across multiple CIM macros with a fixed hardware budget.
Besides, it incorporates grouped convolution to further decrease computing cycles while maintaining near-lossless model accuracy. 
In addition, TetrisG-SDK integrates a validated CIM hardware simulator to provide accurate system-/application-level estimations of latency, area and energy.

Compared to the single-macro VWC-SDK, the proposed framework achieves a speed-up by $1.2\times$, $1.3\times$, and $1.3\times$ for CNN8, GoogLeNet Inception, and DenseNet40 models, respectively.
When deployed on the simulator, it reduces system-level latency and energy by $2.4\times$ and $1.7\times$ for CNN8, $1.3\times$ and $1.2\times$ for Inception, and $1.3\times$ and $1.6\times$ for DenseNet40, respectively. 
When leveraging macro-level parallelism, TetrisG‑SDK reduces the \emph{Energy-Delay-Area-Product (EDAP)} by $70\%$ for CNN8, $68\%$ for Inception, and $36\%$ for DenseNet40 compared to its non-grouped counterpart.
These results manifest that TetrisG-SDK is a promising solution to efficiently mapping convolutional layers on CIM hardware.
\end{abstract}

%
%
\section{Introduction}
\label{sec:intro}
\input{section/intro}
\section{Background}
\input{section/background}
\label{sec:background}

%
%

\section{Proposed Mapping Framework}

\input{section/mapping}

\label{sec:mapping}

%
%

\section{Experimental Result}

\input{section/result.tex}

\label{sec:result}
%

%
%

\section{Conclusion}

\input{section/conclusion}

\label{sec:conclusion}



\bibliographystyle{unsrt}
\bibliography{refs}

\end{document}

%% file: section/intro.tex
In the era of Artificial Intelligence (AI), Convolutional Neural Networks (CNNs) have emerged as a prevalent paradigm across a wide range of AI applications, including image recognition, natural language processing, video analysis \cite{cnn-inception2}\cite{cnn-action}\cite{li2021survey}, etc. CNNs deploy convolution as a fundamental arithmetic operation where inputs are convolved with various filters to extract high-dimensional features for inference purposes. Given the data-intensive nature of CNNs, it is crucial to map convolutional layers along with the associated data onto hardware efficiently so that the computational performance can be maximized \cite{cnn-hardware1}\cite{cnn-hardware4}.


\begin{figure}[h!]
    \centering
    \includegraphics[width=\linewidth, height=4cm]{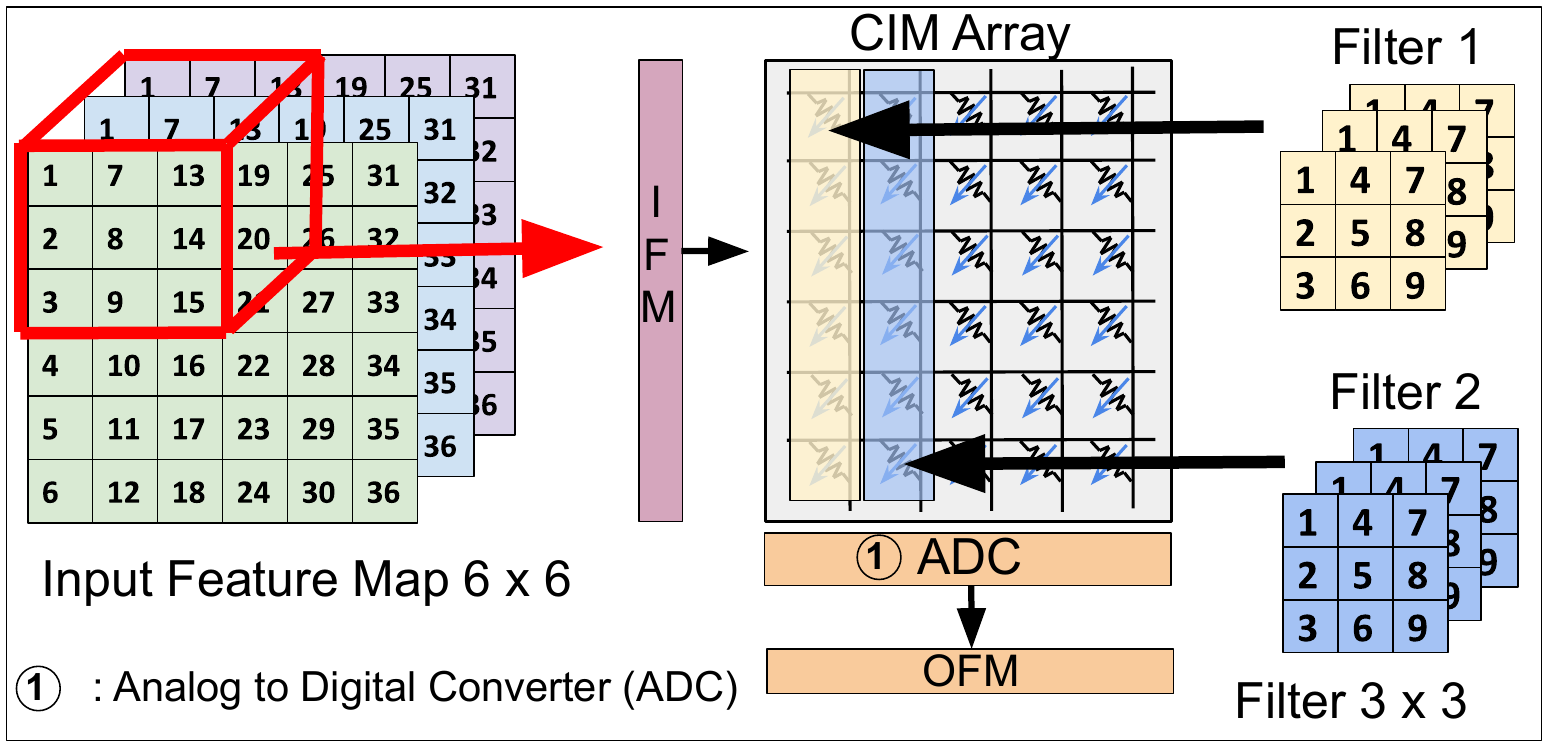}
    \caption{Illustration of convolution operations in a CIM macro where the input features are sent to the input of the array while the filters are mapped onto the bitcells. Adapted from \cite{10558042}.}
    \label{cnn-operation-in-CIM}
\end{figure}

Compute-In-Memory (CIM) technology \cite{cim1}\cite{cim2}\cite{11370184} has gained increasing popularity due to its intrinsic energy efficiency for CNN computations, surpassing conventional von Neumann architectures \cite{Zidan2018}\cite{cim3}\cite{cim4}. The key advantage of CIM stems from its ability to eliminate the memory wall bottleneck in data transfer. By co-localizing computing and data storage within a single memory macro, CIM minimizes the energy consumed during data transfer, resulting in improved overall efficiency. \textcolor{black}{However, state-of-the-art accelerators incorporate many such macros where mapping must not only account for per‑macro efficiency but also how workload is distributed across multiple macros to minimize compute latency. }In this work, we propose a novel convolutional layer mapping methodology for CIM-based CNN acceleration. 
\textcolor{black}{Fig. \ref{cnn-operation-in-CIM} illustrates the architecture of a CIM macro for our framework. Specifically, the kernel weights are mapped onto the memory bitcells, while the Input Feature Maps (IFMs) are loaded to the input wordlines without digital-to-analog converters as we leverage multiple cycles to convert multi-bit input to the wordline voltages one by one as performed in \cite{neurosim}. After accumulating over the CIM crossbar, Output Feature Maps (OFMs) are obtained along the vertical bitlines. 
Given the wide adoption of SRAM-based mixed-signal/analog CIM circuits \cite{Biswas}\cite{XSI19}\cite{dong2020}, our work will primarily discuss the mapping methodology for this type of architecture.}

It is vital to reduce the number of computing cycles and improve array utilization in CIM so that throughput and compute efficiency (i.e., TOPS/W) can be enhanced eventually. 
Several mapping algorithms have been proposed to improve utilization and/or reduce computing cycles. Image to column (img2col)\textcolor{black}{\cite{im2col}} is one of the most common mapping techniques, which unrolls input features to a kernel size. 
Though its array utilization can be high, this mapping cannot reuse input data and thus requires a large number of duplicated IFMs over time, rendering computational latency overhead. 
Shift-and-Duplicate-Kernel (SDK) mapping has been proposed \cite{sdk1}\cite{sdk2} to address the problem by reusing input data within a larger parallel window so that multiple kernel windows can be implemented in parallel. However, the computing cycles are not minimal as they apply a rigid window to the entire channels, which can require more CIM arrays during mapping. 
To address it, Variable-Window SDK (VW-SDK) \cite{vwsdk} has been proposed to further improve the mapping efficiency. 
\textcolor{black}{It partitions the input channels and filter kernels into tiles, leading to fewer computing cycles compared to SDK.}
Despite its advantages, VW-SDK is still suboptimal. \textcolor{black}{Firstly, the current SDK variants only map CNNs onto a single macro, leaving macro‑level parallelism unexplored. Secondly,}
\textcolor{black}{it merely allows for one fixed window shape for the entire layer and doesn't account for the variety of input channel partitions and marginal space at IFMs. This results in inefficient CIM array utilization and extra computing cycles.}
Variable-Window-Channel SDK (VWC-SDK) \cite{vwcsdk} is a variation of VW-SDK, enabling channel-pruning before mapping. Despite its attempt to decrease the computing cycles with pruning, it still suffers from \textcolor{black}{single-macro mapping and} the inherent inefficiency of rigid window shapes.

Tetris-SDK\cite{10558042} is an efficient mapping technique that provides an adaptive set of parallel windows for convolutional layers. The name “Tetris” is inspired by the classic video game, chosen to represent its flexibility and adaptability to varying window sizes. On top of it, we extend the work with optional grouped convolutions to further reduce the computing cycles while the letter "G" emphasizes this update. 
By employing optimal sliding window shapes through adaptation and grouping, 
\textcolor{black}{
TetrisG-SDK enhances CIM utilization of marginal space by $1.6\times$, $2\times$, and $1.5\times$ for CNN8, GoogLeNet Inception, and DenseNet40, respectively compared to VWC-SDK. 
It speeds up convolutional operations by $1.2\times$, $1.3\times$, and $1.3\times$ on CNN8, GoogLeNet Inception, and DenseNet40, respectively.}
\textcolor{black}{Moreover, we integrate the DNN+NeuroSim \cite{neurosim}, a well-established CIM hardware simulation platform to evaluate the performance of TetrisG-SDK with system-level environment
for real-world applications.}
\textcolor{black}{Compared to VWC-SDK, TetrisG-SDK reduces the latency and energy by $2.4\times$ and $1.7\times$ for CNN8, $1.3\times$ and $1.2\times$ for GoogLeNet Inception, and $1.3\times$ and $1.6\times$ for DenseNet40 when deployed on DNN+NeuroSim.} \textcolor{black}{Furthermore, when leveraging macro parallelism, TetrisG‑SDK reduces the \emph{Energy-Delay-Area-Product (EDAP)} by $70\%$ for CNN8, $68\%$ for GoogLeNet‑Inception, and $36\%$ for DenseNet40.}
The contribution of the work is summarized as follows.
\textcolor{black}{\begin{enumerate}
    \item \textcolor{black}{It proposes a macro‑grid search algorithm that selects the best window configuration for multiple CIM macros under a hardware budget and reveals EDAP for in-depth analysis. Our results show that the optimal macro configuration is workload-dependent.}
    \item It introduces a square-inclined window to minimize the required rows of a CIM array while executing the same number of convolutions. As a result, more input channels can be mapped onto the same CIM array.
    \item It optimizes the mapping for marginal space when the sliding window approaches the border of the IFMs, resulting in improved array utilization and a reduced number of windows. It also provides multiple windows to adapt to the input channel depth caused by partitioning. 
    \item It deploys grouped convolutions to optimize convolutional computation on CIM arrays, further decreasing the number of computing cycles with near-lossless accuracy. Implementing grouped convolutions in TetrisG-SDK leads to reduced computational complexity and \textcolor{black}{better hardware scalability as well as }energy efficiency.
\end{enumerate}}
\textcolor{black}{The remaining manuscript is organized as follows. Section II provides the background of this work and a brief review of state-of-the-art SDK mapping algorithms. Section III elaborates on the proposed mapping algorithms. Section IV presents the experimental results and compares them against benchmark mapping methods. Finally, Section V concludes the paper.}

%% file: section/background.tex
\subsection{CIM Architecture Model and Deployment}
\begin{figure}[h!]
    \centering
    \includegraphics[width=\linewidth, height=3cm]{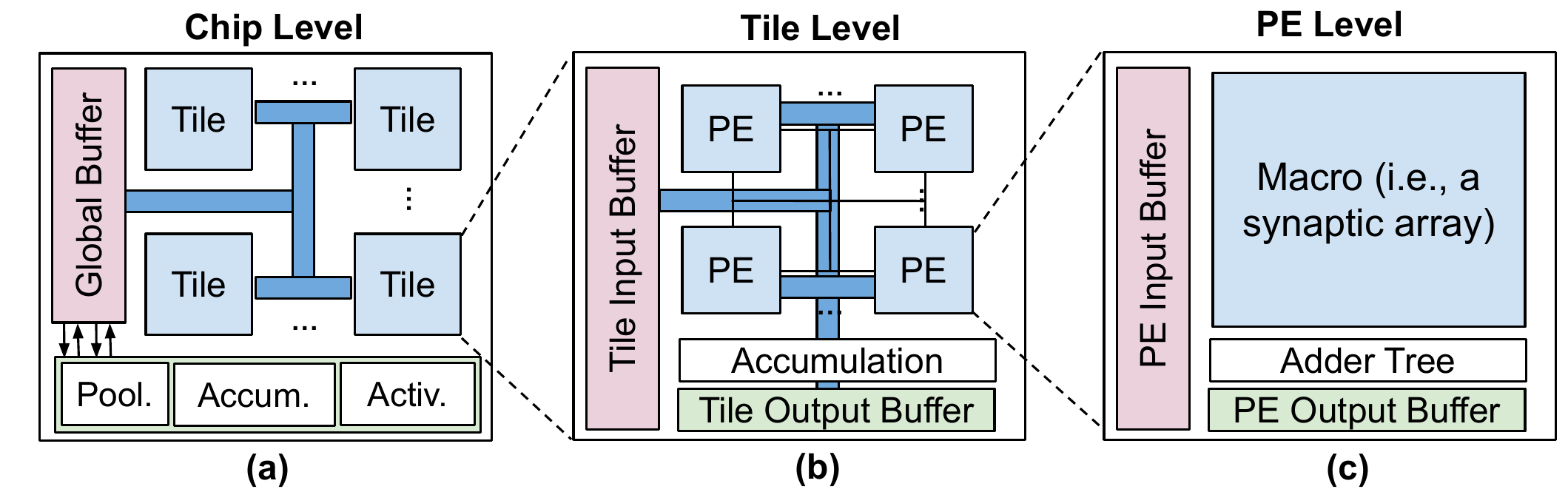}
    \caption{\textcolor{black}{Hardware architecture for CNN accelerator adapted from DNN+NeuroSim\cite{neurosim} where (a) chip level design, (b) tile level design, and (c) PE level design are shown, respectively.}}
    \label{neurosim-chip}
\end{figure}

\begin{figure}[h!]
    \centering
        \includegraphics[scale=0.23]{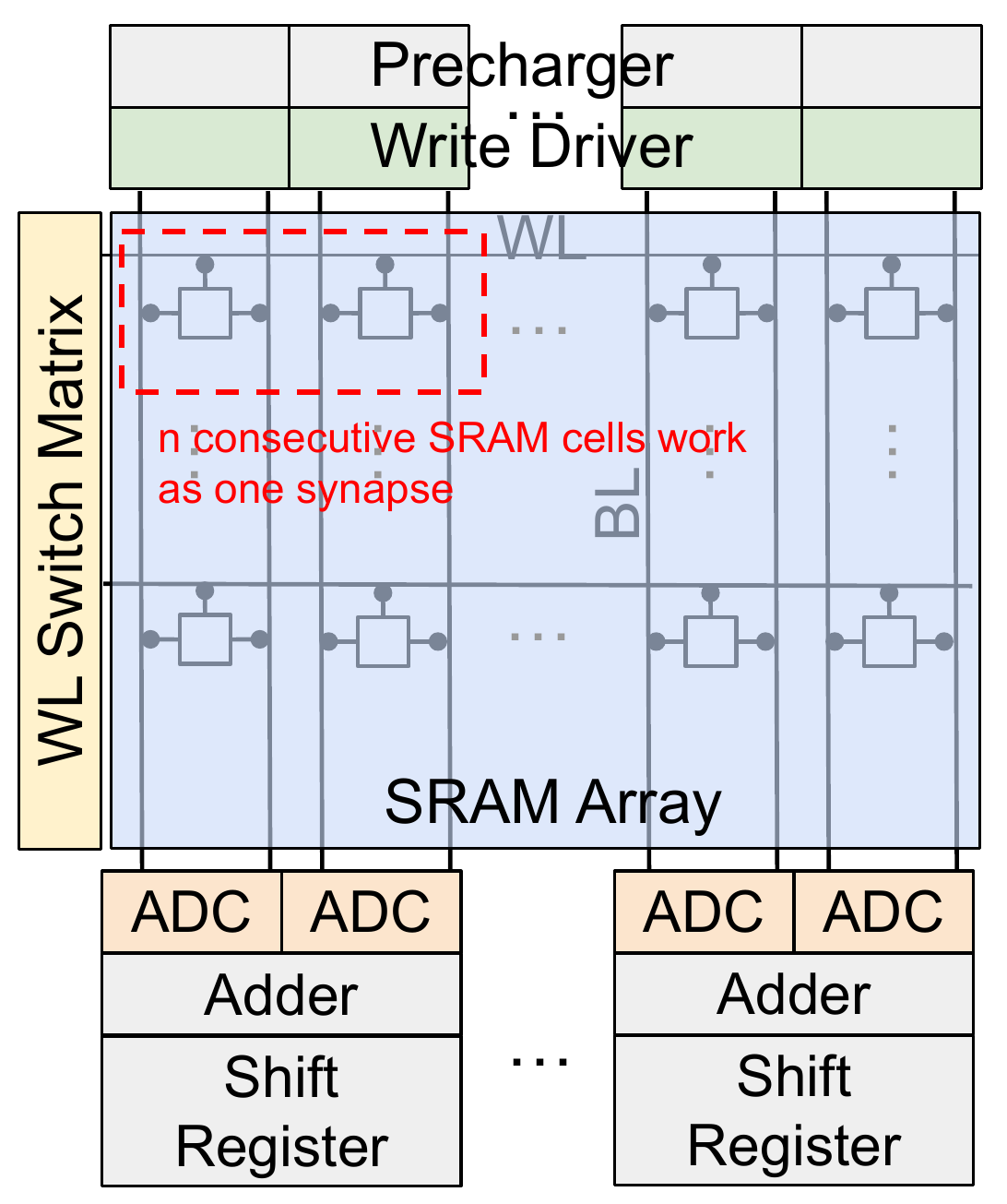}
    \caption{A parallel read-out synaptic sub-array based on SRAM adapted from \cite{neurosim}. Multiple-bit weights are stored across consecutive SRAM cells to work as one synapse.}
    \label{neurosim-sram}
\end{figure}

Fig. \ref{neurosim-chip} depicts the hardware architecture utilized in DNN+NeuroSim \cite{neurosim}, which is further leveraged by TetrisG-SDK for deployment. The top level of the chip consists of tiles, a global buffer, and peripherals including pooling, accumulation, and activations units. In each tile, the architecture is further partitioned into Processing Elements (PEs), accumulation units, and tile-specific input and output buffers. Each PE comprises \textcolor{black}{one CIM macro (i.e., a synaptic array)}, supplemented by adder trees and local buffers \cite{neurosim}. 


To effectively map TetrisG-SDK onto the designated hardware, we firstly determine the chip's floor plan based on the network topology and mapping algorithm. Specifically, traces of synaptic weights and neural activations are unrolled according to the window shape before being saved and sent to the NeuroSim core. Subsequently, these traces are partitioned and allocated to specific chip locations according to the floor plan. The buffer stores all the IFMs and weight kernels. Each tile is designed to accommodate at most a single layer, and further partitioned into PEs to accommodate the weight-matrix \cite{neurosim}. Through the process, the unique window shapes and computation cycles of TetrisG-SDK are utilized to derive final performance metrics, such as latency and dynamic energy consumption. The detailed implementation result are presented in Section IV.

As previously discussed, this paper specifically focuses on mapping onto SRAM-based CIM architectures.
Fig. \ref{neurosim-sram} illustrates the architecture of a CIM macro with SRAM cells at the PE level in DNN+NeuroSim where the synaptic array translates multi-bit weight precision into a series of consecutive 1b SRAM cells. 
In this homogeneous CIM architecture, the IFMs are unrolled to bits and applied sequentially to word lines via the WL switch matrix. Multiplication is performed in the CIM macro. The accumulated partial-sum results will be collected along columns simultaneously with high-precision flash-ADCs. Finally, the adders and shift registers are used to iteratively shift and accumulate partial sums across multiple IFM cycles \cite{neurosim}.

\subsection{Mapping of Convolutional Layers in CIM Arrays}

\begin{figure}[h!]
    \centering
    \includegraphics[width=\linewidth, height=20cm]{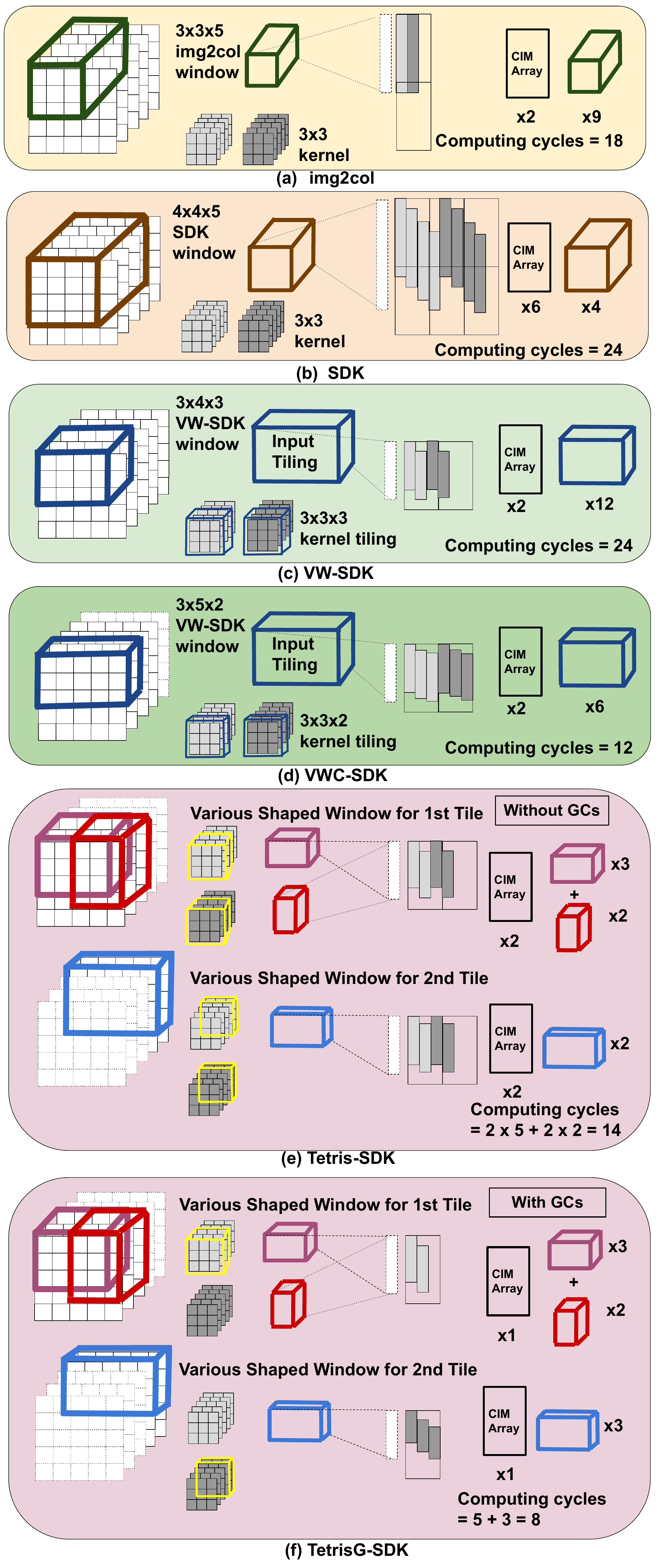}
    \caption{\textcolor{black}{Mapping methods of a convolutional neural network in terms of computing cycles including (a) img2col, (b) SDK, (c) VW-SDK, (d) VWC-SDK, (e) Tetris-SDK and (f) proposed TetrisG-SDK for a single macro with array multiplexing. Adapted from \cite{10558042}.}}
    
    \label{fig:mapping_summary}
\end{figure}

\begin{figure*}[h!]
    \centering
    \includegraphics[width=18cm, height=8cm]{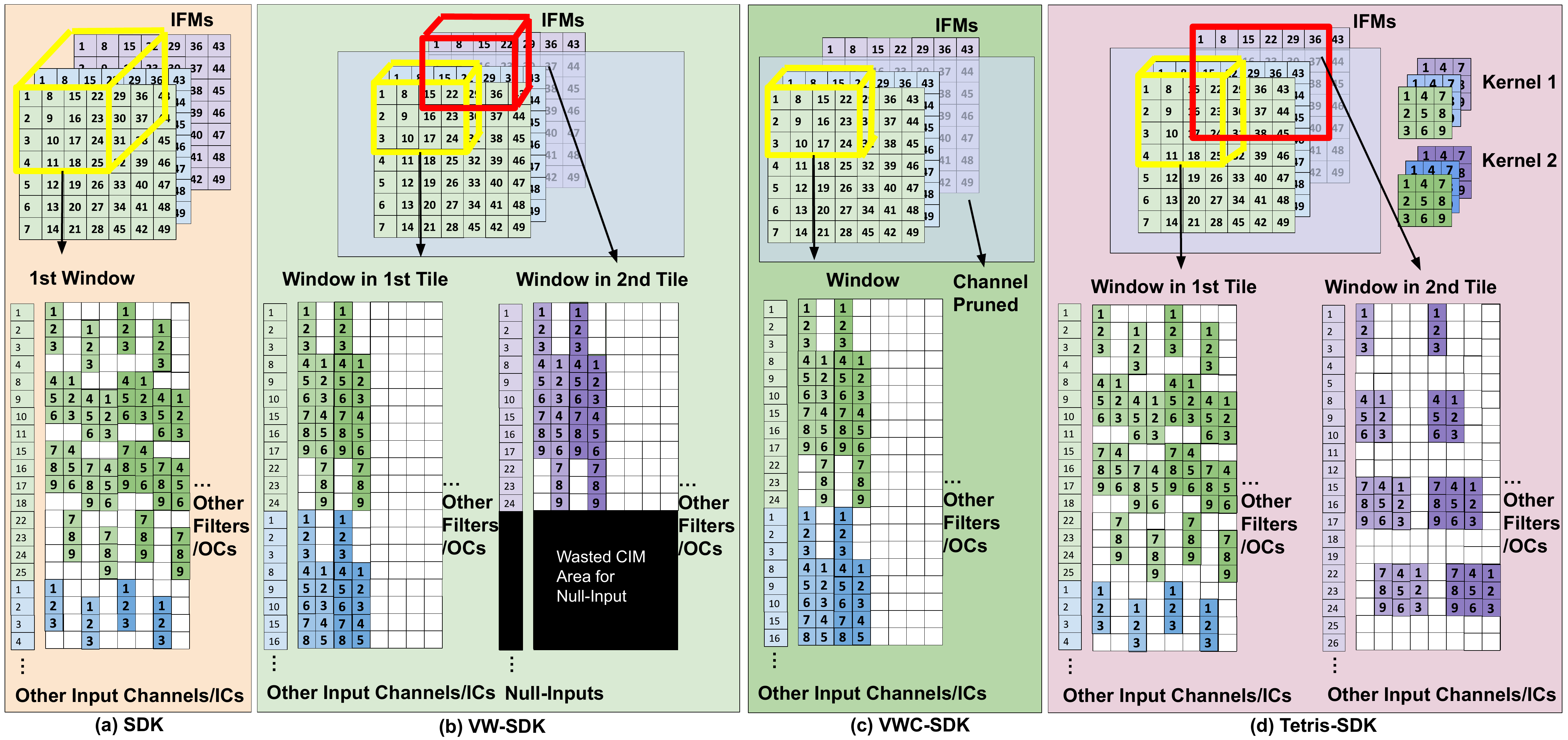}
    \caption{\textcolor{black}{Methods for mapping the convolutional weights into a CIM array, assuming a $3 \times 3$ kernel window for (a) SDK, (b) VW-SDK, (c) VWC-SDK, and (d) Tetris-SDK. The weight kernel matrices remain the same for all mappings. The various colors in the vertical direction of the CIM arrays correspond to different channels in the same weight matrix. The colors in the horizontal direction imply different weight matrices.}}
    \label{fig:background-all-mappings}
\end{figure*}

\begin{figure}[h!]
    \centering
    \includegraphics[width=\linewidth, height=10cm]{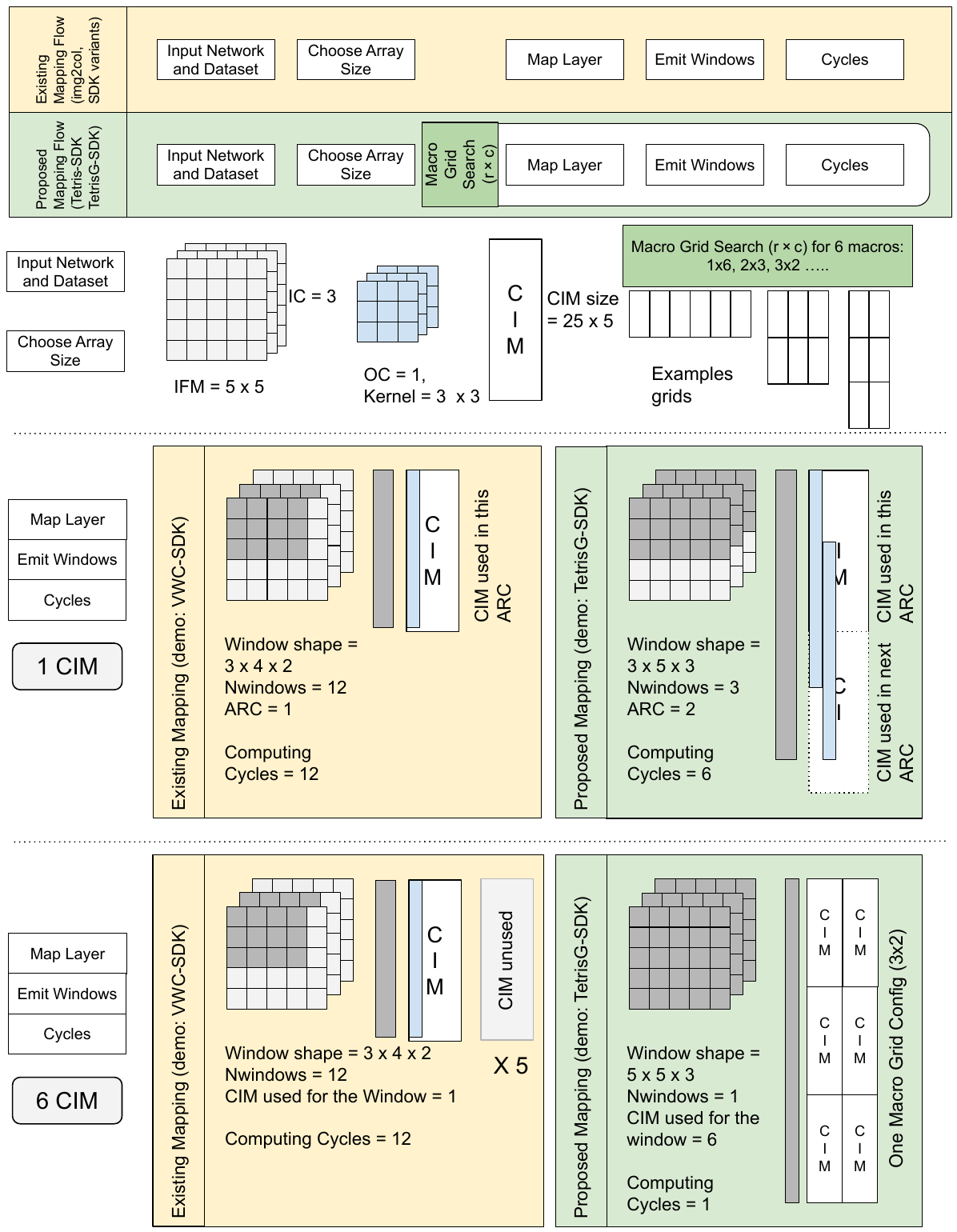}
    \caption{\textcolor{black}{Existing flows (img2col, SDK family) take the array size of a \emph{single} CIM macro as input and map each layer once. The proposed flow adds a \emph{macro‑grid search} that enumerates all \(r{\times}c\) macro arrangements under a fixed budget (e.g., six macros) and selects the arrangement that achieves the lowest total cycle count.}}
    \label{fig:mapping_macro_difference}
\end{figure}

Efficient mapping of IFMs and kernel weights plays a vital role in optimizing the performance of the CIM-based computing. In this context, we delve into several widely adopted mapping algorithms for CIM. For simplicity, we adopt a $40\times15$ CIM array to illustrate how a convolutional layer can be mapped. \textcolor{black}{The 5b weights are mapped horizontally onto the consecutive bitlines in the CIM arrays.} We will derive the calculation of computing cycles for general cases in the subsequent section. 


\textbf{\textit{Image to column (img2col)}:} Fig. \ref{fig:mapping_summary}(a) illustrates img2col\textcolor{black}{\cite{im2col}}, a well-known technique in CPUs and GPUs for convolutional layer mapping. Img2col unrolls the IFMs with respect to the kernel size and duplicates the input as the kernel window slides, enabling the reusing of the weights over time. 
However, this mapping cannot reuse the input features, resulting in a significant number of duplicated IFMs over time with extra computing cycles. 
\textcolor{black}{As Fig. \ref{fig:mapping_summary}(a) depicts, the img2col mapping requires $2$ arrays to fit the $3\times3\times5$ window and $9$ slides to traverse the $5\times5\times5$ IFMs. Hence, the number of computing cycles becomes $18$ (i.e., $2\times9=18$). Moreover, if more CIM arrays are needed due to the increased number of filters, the computing cycles will surge.} \textcolor{black}{A recent work \cite{benchmarking} categorized several variants of img2col mapping, focusing on the traditional unrolling model of convolutions at the PE level. This work primarily benchmarks the variants, particularly on how they manage input data reuse and output data reduction with dedicated hardware. However, the variants always apply the same window size for mapping, suffering from suboptimal computing cycles.}

\textbf{\textit{Shift-and-Duplicate-Kernel (SDK)}:} SDK has been proposed \cite{sdk1}\cite{sdk2} to address the problem by reusing input data within a larger sliding parallel window. The sliding window encompasses multiple kernel windows, allowing for parallel convolutions with less frequent input duplication. 
Although it may lower the computing cycles compared to img2col, 
the computing cycles are sub‑optimal because a rigid window is applied across all channels, requiring a great number of CIM arrays for mapping. In some cases, it even fails to improve the computing cycles when compared to img2col. This is because the fixed larger parallel windows require more CIM arrays, and this drawback outweighs the benefits of multiple kernel window reuse within the parallel windows. As illustrated in Fig. \ref{fig:mapping_summary}(b), a total of $2\times3$ CIM arrays is required to unroll the input parallel windows and output channels, respectively. 
Hence, the number of computing cycles rises to $24$ (i.e., $2\times3\times4 = 24$) as compared to img2col.

\textbf{\textit{Variable Window-SDK (VW-SDK)}} \cite{vwsdk} partitions input channels and examines how different window shapes affect the computing cycles. 
A search algorithm identifies one general window with fewer computing cycles compared to SDK by employing kernel tiling and input tiling. 
However, the sliding window is not optimal in row utilization and fails to incorporate sufficient input channels. Moreover, it does not address the mapping efficiency in the marginal space of the input features where the window is wider than the remaining IFMs. \textcolor{black}{The marginal space refers to the portion of the IFMs that resides on the border and cannot be accommodated by a full sliding window in conventional mapping, resulting in null inputs when mapped to memory arrays.}
Besides, the solution cannot adapt to different channel depths after tiling, as the window is applied to all tiles with the same channel depth. 
\textcolor{black}{Therefore, in some cases, the windows it proposed do not speed up the computing. Fig. \ref{fig:mapping_summary}(c) depicts the mapping based on the VW-SDK algorithm. The computing cycles are $24$ as it requires $2$ CIM arrays and $12$ slides to cover the entire IFMs (i.e., $2\times12=24$), which fails to improve based on img2col.}

\textcolor{black}{\textbf{\textit{Variable Window and Channel-SDK (VWC-SDK)}} \cite{vwcsdk} is an extended work of VW-SDK, allowing for channel pruning to decrease the computing cycles as shown in Fig. \ref{fig:mapping_summary} (d). By removing the residual channels, VWC-SDK further reduces the computing cycle to $12$. However, this technique only works for selected layers, compromising its scalability and efficacy \cite{vwccode}.}
\textcolor{black}{Thus far, all discussed SDK variants optimize mapping with respect to one CIM macro.  
They do not exploit the multiple macros that exist in modern accelerators.}

\textcolor{black}{\textbf{\textit{Tetris-SDK}} is a more adaptive mapping algorithm. It searches for square-inclined windows, adapts to the borders of IFMs, and utilizes different channel partitions to lower the overall computing cycles and increase the utilization of CIM arrays. 
Consequently, it allows more efficient convolutions in CIM-based architectures with its adaptive mapping. 
As illustrated in Fig. \ref{fig:mapping_summary} (e), Tetris-SDK can achieve performance improvement even when SDK and VW-SDK fail to outperform img2col. 
Specifically, it generates $3$ parallel windows and $2$ marginal windows to cover the first tile of IFMs and $2$ parallel windows to slide for the second tile. As both the first tile and second tile require $2$ CIM arrays, the total computing cycles are $14$ (i.e., $(3 + 2 + 2)\times2 = 14$). }

\textcolor{black}{Fig. \ref{fig:background-all-mappings} further illustrates the difference in terms of array utilization between SDK, VW-SDK, VWC-SDK, and Tetris-SDK \textcolor{black}{under single macro}. As shown in Fig. \ref{fig:background-all-mappings}(a), SDK adopts a rigid and fixed shape across all channels. VW-SDK in Fig. \ref{fig:background-all-mappings}(b) partitions the channels into different tiles. However, the same window is applied to all tiles regardless of channel partitions and the null inputs are incorporated into the second tile as the remaining channel depth is less than the window depth. This leads to wasted area and low utilization of the CIM array. Fig. \ref{fig:background-all-mappings}(c) illustrates the residual channel pruning proposed by VWC-SDK to account for the depth variability. However, this approach only applies to selective layers in VWC-SDK. Fig. \ref{fig:background-all-mappings}(d) shows that Tetris-SDK removes the null-inputs in (b) by adaptive windows based on channel partitions. It allows more efficient convolutions in CIM array architectures with well-optimized mapping.}

\textcolor{black}{\textbf{\textit{TetrisG-SDK}} further enhances efficiency by allowing grouped convolutions as illustrated in Fig. \ref{fig:mapping_summary}(f). \textcolor{black}{When the group number becomes $2$, the computing cycles decrease to $8$ as fewer CIM arrays are required to unroll the parameters. With grouped convolutions, the window shape can morph to a larger one, accommodating more kernel windows within the parallel windows and further increasing the CIM array utilization.}} 
\textcolor{black}{Moreover, TetrisG-SDK exhibits a key advantage over prior works by systematically exploring the possible macro grid configurations with the given hardware budget. Specifically, the search engine pinpoints the optimal arrangement that minimizes the computing cycles as illustrated in Fig. \ref{fig:mapping_macro_difference} where the baseline VWC-SDK consumes 12 cycles while TetrisG-SDK accomplishes in a single cycle through its optimized $3{\times}2$ macro-grid.}

\textcolor{black}{\subsection{Grouped Convolutions}}
\textcolor{black}{Here we introduce group \textcolor{black}{convolutions which can be utilized to further optimize the proposed mapping methodology.}
Grouped convolutions divide conventional convolutions into a set of smaller, parallel convolutions. 
Standard convolutions perform Multiple-Accumulate (MAC) operations on the entire IFMs with kernel filters to produce complete Output Feature Maps (OFMs).
Grouped convolutions split both IFMs and filters into an equal number of groups and perform standard convolutions within each group. The output of each group is concatenated to form the OFMs. This approach enables the network to learn distinct sets of low-level and high-level features \cite{resnext}.
Grouped convolutions are initially proposed in AlexNet \cite{alexnet} to overcome hardware restrictions by distributing the network among multiple GPUs due to insufficient GPU memory.
Subsequent studies such as ResNext \cite{resnext}, Dynamic Group Convolution (DGC) \cite{dgc}, and
Fully Learnable Group Convolution \cite{flgc} have demonstrated that grouped convolutions not only reduce the number of parameters but can also enhance classification accuracy. 
This improvement arises from the unique representation of the IFMs that each filter group learns (i.e., cardinality dimensions), which is less common in conventional convolutions where filter kernels typically correlate with one another \cite{resnext}. 
Grouped convolutions are applicable across various networks and diverse datasets, leading to better learning performance and reduced computational cost.}

\begin{figure}[h!]
    \centering
    \includegraphics[scale = 0.165]{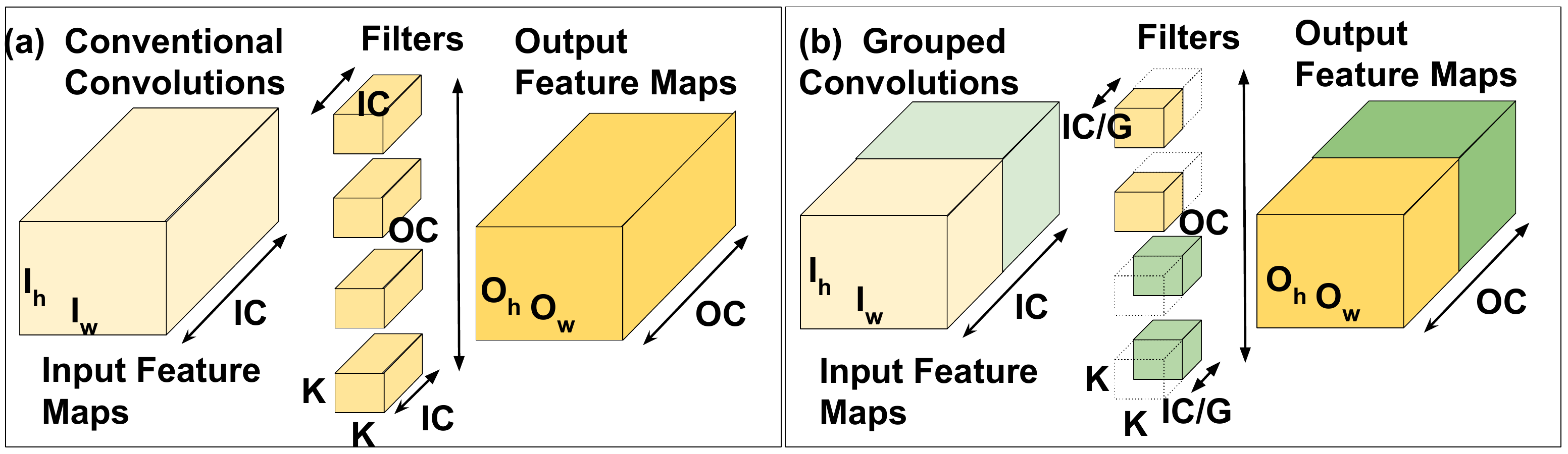}
    \caption{Illustration of conventional convolutions and grouped convolutions.}
    \label{gconv}
\end{figure}

\textcolor{black}{As shown in Fig. \ref{gconv}(a), in conventional convolutions,  the IFMs have a size of $I_h \times I_w \times IC$, where $I_h$, $I_w$, and $IC$ are IFMs height, width, and number of input channels. Each filter has a size of $K \times K \times IC$, and there are a total of $OC$ number of the filters where $K$ represents the height and width of filters and $OC$ accounts for the number of output channels. Hence, the OFMs have a size of $O_h \times O_w \times OC$, where $O_h$ and $O_w$ represent the OFMs height and width, respectively.}
\textcolor{black}{The number of parameters ($N_{params}$) and the number of computations required ($N_{op}$) in conventional convolutions are calculated as the following:}

\begin{equation}
\begin{aligned}
\textcolor{black}{N_{params} = K \times K \times IC \times OC}
\end{aligned}
\end{equation}

\begin{equation}
\begin{aligned}
\textcolor{black}{N_{op} = N_{params} \times O_h \times O_w}
\end{aligned}
\end{equation}


\textcolor{black}{In grouped convolutions, dividing the filters and IFMs is crucial for reducing network complexity and speeding up neural network training. As illustrated in Fig. \ref{gconv}(b), both the IFMs and filters are divided into two groups. Parallel convolutions are performed concurrently on their respective group, and each filter possesses precisely half the number of parameters compared to a standard convolutional layer shown in Fig. \ref{gconv}(a). Subsequently, the OFMs of an individual group are concatenated, halving the number of computations in this example.}
\textcolor{black}{Generally, suppose the number of groups is $G$, the number of parameters ($NG_{params}$) and the number of computations required ($NG_{op}$) are reduced to $\frac{1}{G}$ in grouped convolutions, and can be calculated as follows:}

\begin{equation}
\begin{aligned}
\textcolor{black}{NG_{params} = K \times K \times \frac{IC}{G} \times \frac{OC}{G} \times G}
\end{aligned}
\end{equation}

\begin{equation}
\begin{aligned}
\textcolor{black}{NG_{op} = NG_{params} \times O_h \times O_w }
\end{aligned}
\end{equation}


\subsection{Calculation of Computing Cycles and Array Utilization}
\begin{figure}[h!]
    \centering
    \includegraphics[width=\linewidth, height=10cm]{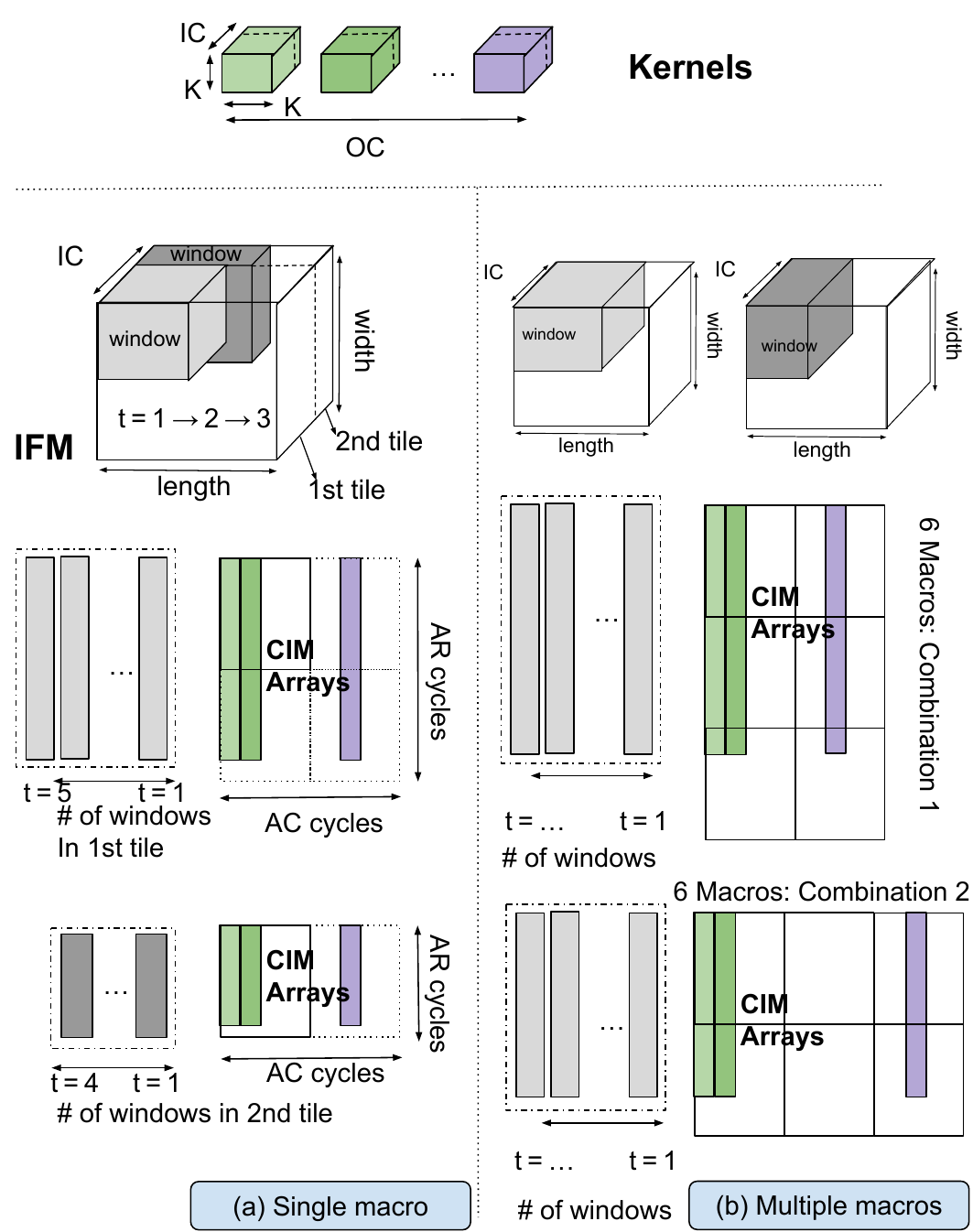}
    \caption{\textcolor{black}{Illustration of computing‑cycle calculation for (a) single‑macro execution and (b) distribution over a $6$ macro grid with different possible combinations. Adapted from \cite{10558042}.}}
    \label{compute-cc}
\end{figure}

\textcolor{black}{As discussed in Section I, we adopt computing cycles and array utilization which are widely used in recent mapping frameworks \cite{vwsdk}\cite{vwcsdk}\cite{Park2024KARS} to assess the efficiency of the mapping process. The computing cycle is the required number of clock cycles for computation from the presence of effective input data to the generation of OFMs.
As depicted in Fig. \ref{compute-cc}, a CIM array has fixed Array Rows ($AR$) and Array Columns ($AC$). The Array Row Cycles ($AR_c$) refer to the cycles needed to unroll the IFMs window vertically \textcolor{black}{using a single macro}. The Array Column Cycles ($AC_c$) are defined as the cycles required to unroll the IFMs window horizontally \textcolor{black}{using a single macro} for the partial sums of the OFMs.}
\textcolor{black}{When \emph{multiple} macros are available, these per‑macro cycles are overlapped in time. $AR_c$ and $AC_c$ are executed in parallel.}

\begin{figure}[h!]
    \centering
    \includegraphics[width=\linewidth, height=4cm]{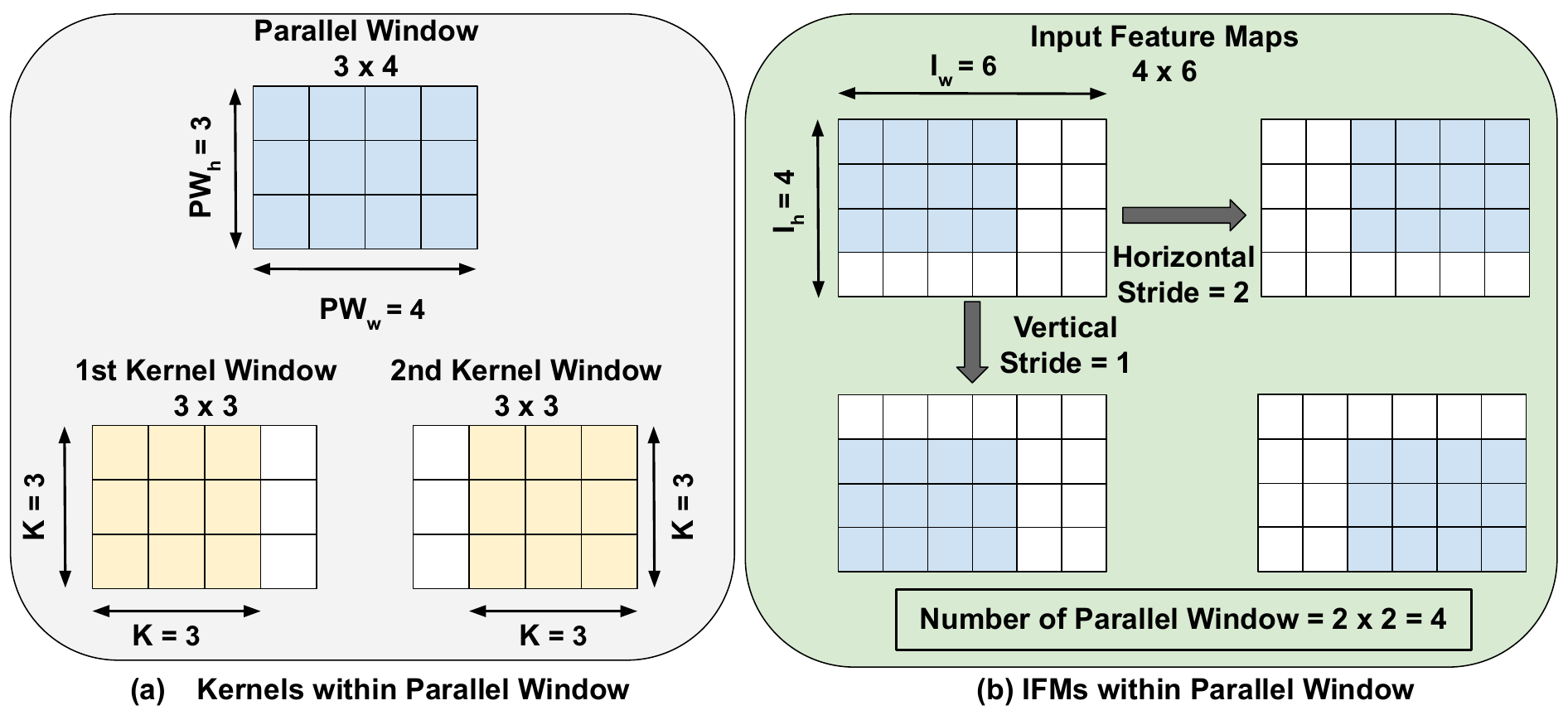}
    \caption{Illustration of (a) kernel computations within a parallel window and (b) IFMs with parallel windows. Adapted from \cite{10558042}.}
    \label{stride}
\end{figure}

\textcolor{black}{Computing cycles on a single macro ($CC_{\text{single}}$) describe the time to finish a layer with one CIM array. They are determined by the number of parallel windows ($N_{windows}$) required in a specific tile $i$, $AR_c$, $AC_c$, and the total number of partitions (i.e., $N$) in Equation \ref{eq:cc_single}. 
}


\begin{equation}
CC_{\text{single}}
      =\sum_{i=1}^{N} N_{\text{windows}}(i)\,AR_c(i)\,AC_c(i)
\label{eq:cc_single}
\end{equation}

\textcolor{black}{When the same layer is mapped onto a grid of $P$ macros (e.g.\ a $2{\times}3$ or $3{\times}2$ grid in Fig.\,\ref{compute-cc}(b)), both the \emph{shape} of each window and the \emph{$N_{\text{windows}}$} of windows change. After the outer search (Section III) selects the best macro grid, the final computing cycles on multiple macros ($CC_{\text{multi}}$) with $P$ macros are therefore reduced without the need to consider $AR_c$ and $AC_c$. $N^{(P)}_{\text{windows}}$ are re‑computed after the window set is resized for a $P$-macro grid as:}

\begin{equation}
CC_{\text{multi}}
      =\sum_{i=1}^{N} N^{(P)}_{\text{windows}}(i)
\label{eq:cc_multi}
\end{equation}

\textcolor{black}{Fig. \ref{stride} illustrates the concept of parallel windows and the calculation of $N_{windows}$. Fig. \ref{stride}(a) depicts the kernel computation within a parallel window. Supposing with a parallel window of $3 \times 4$ (i.e., $PW_h\times PW_w$), $2$ kernel windows with a size of $3 \times 3$ (i.e., $K \times K$) are embedded for the computation.}
\textcolor{black}{In Fig. \ref{stride}(b), we assume an IFM with a size of $4\times6$ (i.e., $I_h \times I_w$) where the shapes of the parallel windows and the filter kernel are unchanged. Without counting the marginal windows, the horizontal stride is $4 - 3 + 1 = 2$ (i.e., $PW_w - K + 1$). It needs $\lfloor\frac{6 - 4}{2}\rfloor + 1 = 2$ horizontal windows (i.e., $N_{horizontal}$) to cover the IFM, (i.e., $\lfloor\frac{I_w -PW_w}{PW_w - K + 1}\rfloor + 1$ ). Similarly, the IFM needs $\lfloor\frac{I_h -PW_h}{PW_h - K + 1}\rfloor + 1$ windows vertically (i.e., $N_{vertical}$). Hence, as Equation \ref{eq:Nwindow} shows, the total number of parallel windows ($N_{windows}$) is determined by a product of the two factors and then summed with the number of marginal windows (i.e., $N_{marginal}$). A more detailed explanation of the calculation of $N_{marginal}$ will be provided in Algorithm \ref{alg:marginal}.}


\begin{equation}
\begin{aligned}
N_{windows} 
= {} & N_{horizontal} \times N_{vertical} + N_{marginal} \\
= {} & (\lfloor\frac{I_w -PW_w}{PW_w - K + 1}\rfloor + 1) \\  
\times & (\lfloor\frac{I_h -PW_h}{PW_h - K + 1}\rfloor + 1) \\ 
+ & N_{marginal}
\end{aligned}
\label{eq:Nwindow}
\end{equation}




\textcolor{black}{CIM array utilization is computed as the ratio of CIM cells mapped with weights ($WC$) to the total number of CIM cells, which is defined as the product of $AR$ and $AC$ (Equation \ref{eq:util}). This metric serves as an indicator of how efficiently the mapping algorithm utilizes the available resources of a CIM array.}

\begin{equation}
\begin{aligned}
Utilization = \frac{WC}{AR \times AC} \times 100 \%
\end{aligned}
\label{eq:util}
\end{equation}

\textcolor{black}{By considering both computing cycles and array utilization, we can effectively evaluate and compare the performance of different mapping algorithms in terms of their efficiency and resource utilization. }

%% file: section/mapping.tex
\subsection{Overall TetrisG-SDK Mapping Algorithm}
\begin{figure}[h!]
    \centering
    \includegraphics[width=\linewidth, height=10cm]{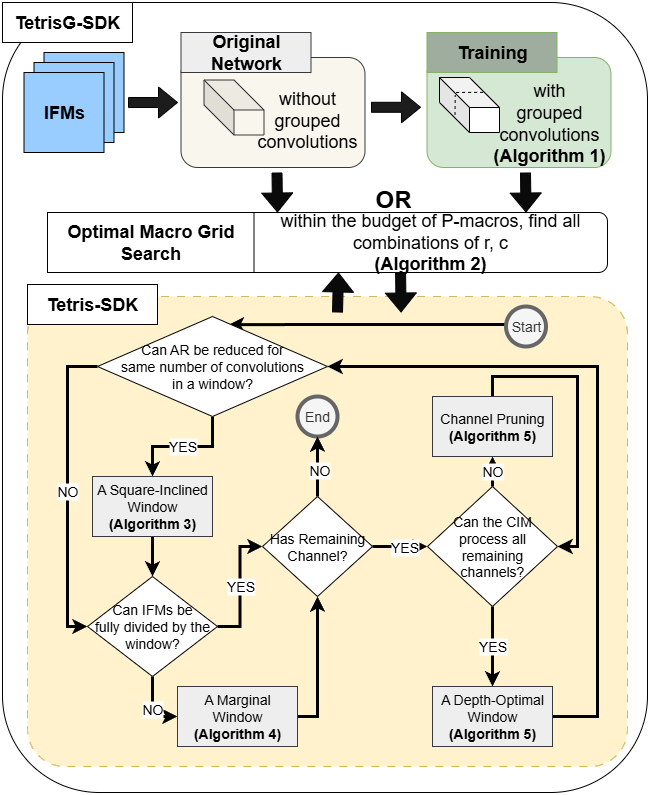}
    \caption{Flowchart of the overall TetrisG-SDK algorithm including the enabling of grouped convolutions, selecting the optimal macro grid, and searching for a square-inclined window, a marginal window, and a depth-optimal window.} 
    \label{flowchart}
\end{figure}

\textcolor{black}{The overall TetrisG-SDK framework consists of the Tetris-SDK algorithms and a training process to implement group convolutions (Fig. \ref{flowchart}). Specifically, Algorithms 1, 2, 3, 4, and 5 are executed to enable/disable grouped convolutions, search for possible macro configurations, square-inclined windows, marginal windows, and depth-optimal windows, respectively. The procedure is described as follows.}

\begin{enumerate}
  \item \textcolor{black}{Train the original network with grouped convolutions until the target accuracy is achieved (i.e., Alg.~\ref{alg:gc}). The process generates the parameters used for inference on CIM-based architectures. If accuracy is significantly degraded, grouped convolutions can be disabled. }
  \item \textcolor{black}{Run the Tetris-SDK algorithms to determine the proposed parallel windows.}
  \begin{enumerate}
  \item \textcolor{black}{The framework first run macro-grid search to execute Alg.~\ref{alg:macro} to select the
        $(r^{\star},c^{\star})$ grid that minimizes $CC_{\text{multi}}$.}
         \item It then determines whether there is a square-inclined window that can increase CIM AR utilization (i.e., Alg.~\ref{alg:sq}). 
         \item Subsequently, it checks the marginal space at the IFMs borders to find out whether a marginal window is necessary (i.e., Alg.~\ref{alg:marginal}).
         \item Tetris-SDK then determines the channel partition and stops when all the channels have been processed. If channels are split into tiles with various depths, it reruns (i.e., Alg.~\ref{alg:sq} and \ref{alg:marginal}) to process other partitions. In the case of pruning, re-training of the models is required to validate the accuracy (i.e., Alg.~\ref{alg:dep}). 
       \end{enumerate}
   \item Finally, a set of optimized windows is obtained for CIM-based architectures. 
\end{enumerate}

\textcolor{black}{\subsection{TetrisG-SDK with Grouped Convolutions}}

\begin{algorithm}[hbt!]
\fontsize{7.5}{10}\selectfont
\caption{\textcolor{black}{TetrisG-SDK with Grouped Convolutions Algorithm}}\label{alg:gc}
Train the original network with grouped convolutions and obtain the group number $G$\;
\KwData{$I$, $K$, $IC$, $OC$, $AR$, $AC$, $G$}
\KwResult{TetrisG-Parallel-Window ($PW$), Computing\_Cycles ($CC$)}
$IC \gets IC/G$\;
$OC \gets OC/G$\;
$PW, IC_t, OC_t \gets K, 0, 0$;
$CC \gets 0$\;
Calculate benchmark CC ($BCC$) with $PW_w \gets K, PW_h \gets K$\;

\For{$PW_w$ in range $(K,I_w)$ and $PW_h$ in range $(K,I_h)$}
{
$IC_t \gets \lfloor(AR /(PW_w \times PW_h))\rfloor$\;
$OC_t \gets \lfloor(AC / ((PW_w-K+1) \times (PW_h-K+1)))\rfloor$\;
$AR_c = \lceil(IC / IC_t)\rceil$\;
$AC_c = \lceil(OC / OC_t)\rceil$\;
Calculate $CC$ for the combination\;
\If{$CC < BCC$}{
      $BCC \gets CC$\;
      \Return{$CC$}\;
    }
}
Run Algorithm 2, 3, 4, and 5 to optimize the found window\;
\Return{$PW, CC$}\;
\end{algorithm}

\textcolor{black}{
Implementing grouped convolutions in TetrisG-SDK is beneficial. Firstly, the reduced computational complexity enables faster processing time. By dividing the input channels into groups, the number of MAC operations is reduced, allowing for more efficient computing and shorter latency. Secondly, by carefully selecting the number of groups and optimizing the mapping, we can increase the hardware resource efficiency while maintaining or even improving the model accuracy. 
Thirdly, the implementation of grouped convolutions improves hardware scalability. By reducing the number of parameters and computations, we can accommodate larger networks with the same footprint, further enhancing the system's overall capabilities.
Moreover, the adoption of grouped convolutions within the TetrisG-SDK framework can contribute to a more flexible and adaptive CIM-based system. As the number of groups can be adjusted to meet specific application requirements or hardware constraints, we can tailor the model to achieve the optimal trade-off between performance, accuracy, and resource usage.}

\textcolor{black}{Grouped convolution and the adaptive-window search in TetrisG-SDK are intrinsically coupled through the CIM macro constraints. Given a grouping factor of $G$, a layer with  a dimension of $(IC,OC)$ is transformed into per-group dimensions as:
\begin{equation}
IC_g = IC/G,\quad OC_g = OC/G
\end{equation}}
\textcolor{black}{
Under a fixed CIM macro size of $(AR,AC)$, the feasibility and efficiency of a parallel window $(PW_h \times PW_w)$ are jointly constrained by both array dimensions. Specifically, $AR$ determines how many input channels can be mapped in parallel for a given window footprint (Equation \ref{eq:AR_constraint}), while $AC$ limits the number of output channels and the parallel kernels within the window that can be processed concurrently (Equation \ref{eq:AC_constraint}). Accordingly, increasing G reduces $(IC_g,OC_g)$, which typically alleviates the constraint imposed by $AR$ and/or $AC$, thereby enlarging the feasible search space of parallel-window configurations. It is worthwhile noting that the adaptive-window search procedure itself (Algorithms 2-5) remains unchanged. The subsequent window and macro optimization steps simply operate on these per-group dimensions.}

\begin{flalign}
\textbf{AR constraint:}\quad & IC_t \times PW_h \times PW_w \le AR && \label{eq:AR_constraint}
\end{flalign}
\begin{equation}
\textbf{AC constraint:}\quad O C_t \times (P W_h - K + 1)\times (P W_w - K + 1) \le AC
\label{eq:AC_constraint}
\end{equation}

\begin{figure}[h!]
    \includegraphics[width=\linewidth, height=10cm]{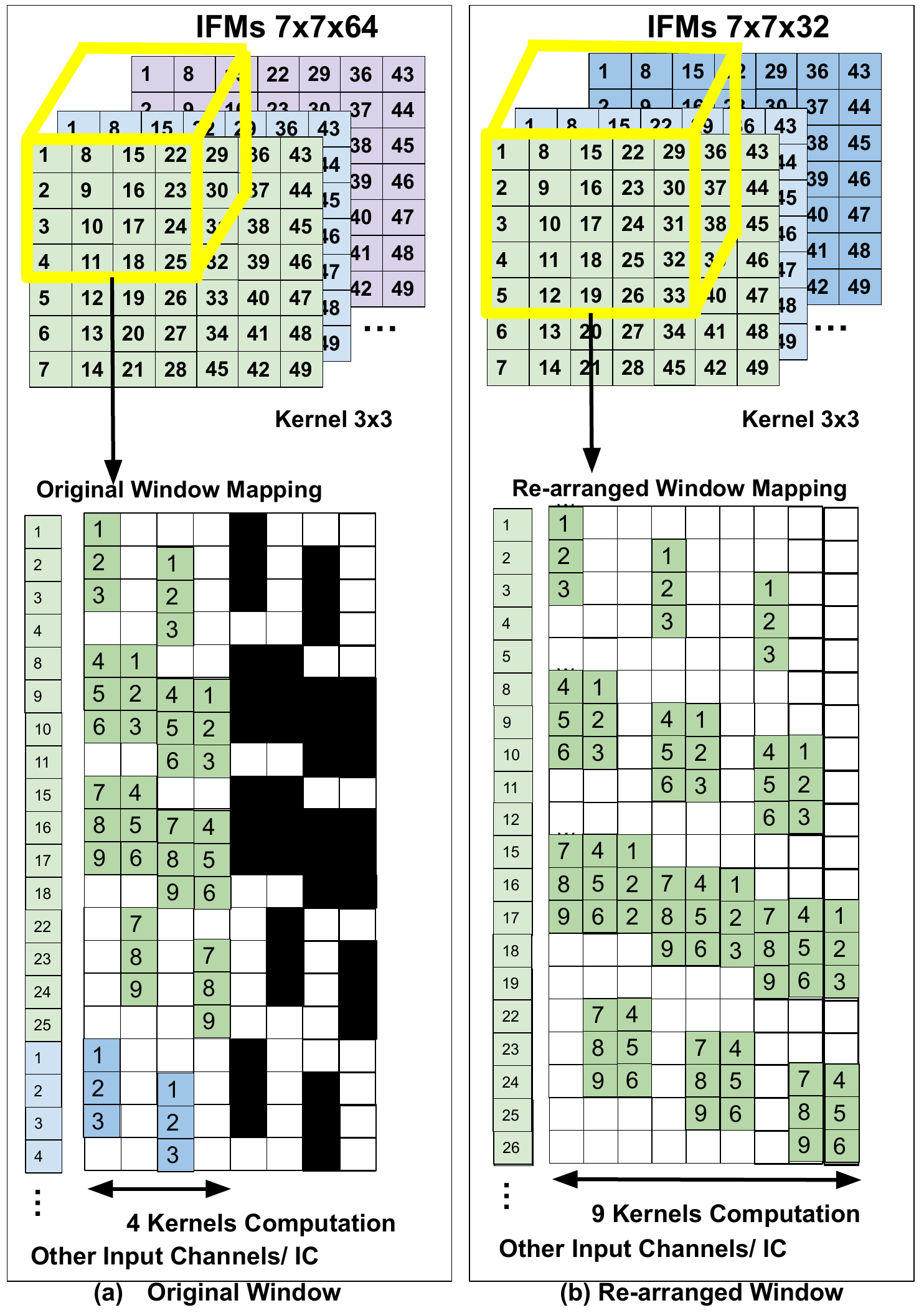}
    \caption{Mapping for TetrisG-SDK with grouped convolutions for (a) original window which accommodates 4 kernel computations and (b) re-arranged window which accommodates 9 kernel computations.}

    \label{fig:sub3 group}
\end{figure}

\textcolor{black}{To incorporate grouped convolutions, we need to determine the number of groups to be applied to the original network using a heuristic approach. First, we ensure that the accuracy reduction is minimal (e.g., less than 0.5\%). Next, we evaluate the speed-up in computing cycles for each group number, incrementally increasing the number of groups and repeating the measurements. We then compare the results across different group numbers to identify the most effective configuration. 
}
Following this approach, we employ \textbf{Algorithm \ref{alg:gc}} to obtain the most suitable mapping strategy, which includes the optimal window shape and the minimal computing cycles for the given network. 
\textcolor{black}{As Fig. \ref{fig:sub3 group} shows, grouped convolutions partitions the IFMs.}
Therefore, it allows the window size to expand if the AC is the bottleneck. More kernels can be extracted and computed from the larger window, which enhances the CIM array utilization.

\subsection{Macro‑Configuration Search}
\textcolor{black}{Given a hardware budget of $P_{\max}$ CIM macros, we enumerate every rectangular grid $(r,c)$ while satisfying $r\!\times\!c\le P_{\max}$. For each candidate grid, we invoke Algorithms~3–5 to obtain the aggregate multi‑macro cycle count $CC_{\text{multi}}(r,c)$ in Equation ~\eqref{eq:cc_multi}.}
\textcolor{black}{Particularly, \textbf{Algorithm~2} is executed offline during the mapping configuration stage and does not consume CIM inference time. It enumerates $\sum_{r=1}^{P_{\max}}\lfloor P_{\max}/r \rfloor = O(P_{\max}\log P_{\max})$ candidate grids $(r,c)$. For practical macro budgets, the search completes in the sub-second range, and therefore does not affect runtime performance.}

\begin{algorithm}[hbt!]
\fontsize{7.5}{10}\selectfont
\caption{Macro‑Configuration Search}
\label{alg:macro}
\KwIn{$P_{\max}$, network topology, Algorithms 1–4}
\KwOut{Best grid $(r^{\star},c^{\star})$, minimal $CC_{\text{multi}}$}

$CC_{\text{best}}\leftarrow\infty$;\;
\For{$r\gets1$ \KwTo $P_{\max}$}{
  \For{$c\gets1$ \KwTo $\lfloor P_{\max}/r\rfloor$}{
    Apply Algs.~1–4 assuming $r{\times}c$ parallel macros;\;
    \If{$CC_{\text{multi}}(r,c)<CC_{\text{best}}$}{
      $CC_{\text{best}}\leftarrow CC_{\text{multi}}(r,c)$\;
      $(r^{\star},c^{\star})\leftarrow(r,c)$\;
    }
  }
}
\Return{$(r^{\star},c^{\star}),\,CC_{\text{best}}$}
\end{algorithm}

\subsection{A Square-Inclined Window}


\textbf{Algorithm \ref{alg:sq}} is proposed to find a square-inclined window as this type of window uses less $AR$ compared to a rectangular window that executes the same amount of convolutions. Consequently, more input channels can be mapped to the same CIM array. 
\textcolor{black}{
Specifically, let the mapping window cover an OFM size of $a \times b$,
which yields $N_{\mathrm{conv}} = a \times b$ convolutions.
For a $K \times K$ kernel, let $t=K-1$, so that the input patch loaded into the macro has a spatial footprint of $A\_in = (a + t)(b + t)= N_{\mathrm{conv}} + t(a+b) + t^2$. 
For fixed $N_{\mathrm{conv}}$ and $K$, minimizing $A_{\mathrm{in}}$ is equivalent to minimizing $(a+b)$. According to the AM-GM Inequality \cite{HardyLP1952}, $a+b \ge 2\sqrt{ab} = 2\sqrt{N_{\mathrm{conv}}}$ with equality where $a=b$.
Thus, the minimum $A_{\mathrm{in}}$ is achieved when $(a,b)$ is as balanced as possible (i.e., the pair closest to $\sqrt{N_{\mathrm{conv}}}$), illustrating the rationale for the near-square factor-pair search in Algorithm 3.}

\begin{algorithm}[hbt!]
\fontsize{7.5}{10}\selectfont
\caption{Finding Square Optimized Window}\label{alg:sq}
\KwData{$I$, $K$, $IC$, $OC$, $AR$, $AC$, $PW_w \times PW_h \times IC_t$, $N_{conv}$}
\KwResult{Square-optimized-window ($SW$), Processed\_channel ($PC$)}
$N_{AR} \gets PW_w \times PW_h \times IC_t$\;
$N_{optimalAR} \gets N_{AR}$\;
$Found\_SO \gets False$\;
  factorize $N_{conv}$\ using square-root\;
  \For {factor $(a, b)$ of $N_{conv}$}
   {\If{$\min((a + K -1)\times(b + K -1)) \leq PW_w \times PW_h$}{
      $SW_w \gets a$\;
      $SW_h \gets b$\;
      $PC \gets \lfloor(\frac{AR}{a\times b})\rfloor$\;
      $N_{optimalAR} \gets a \times b \times PC$\;
      $Found\_SO \gets True$\;
      
    }\Else{$SW_w \gets PW_w$\;
      $SW_h \gets PW_h$\;}
    \Return{$SW, PC$}\;
  
} 
\end{algorithm}

\begin{figure}[h!]
    \centering
    \includegraphics[width=\linewidth, height=12cm]{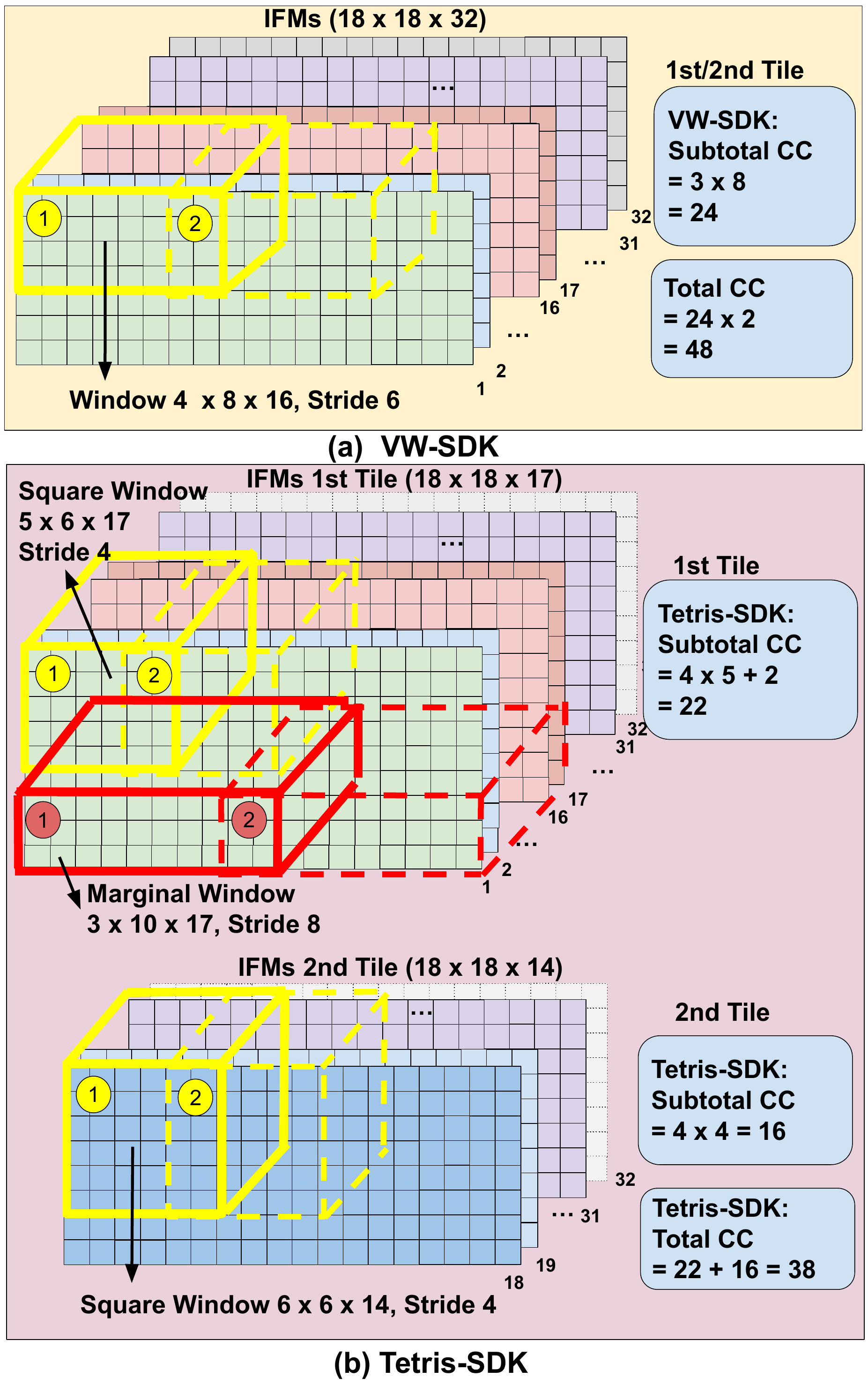}
    \caption{\textcolor{black}{Mapping of CNN8-3 layer in terms of computing cycles (CC) by (a) VW-SDK and (b) Tetris-SDK. Adapted from \cite{10558042}.} 
    }
    \label{fig:cnn8-3}
\end{figure}

\subsection{A Marginal Window}
As described, VW/VWC-SDK can result in inefficient mapping at IFMs borders.
\textbf{Algorithm \ref{alg:marginal}} creates special marginal windows to eliminate the null input in the VW/VWC-SDK window, which is caused when the size of IFMs is not fully divisible by the size of the window. As a result, it enhances CIM utilization and decreases the computing cycles.

\begin{figure}[h!]
    \centering
    \includegraphics[width=\linewidth, height=8cm]{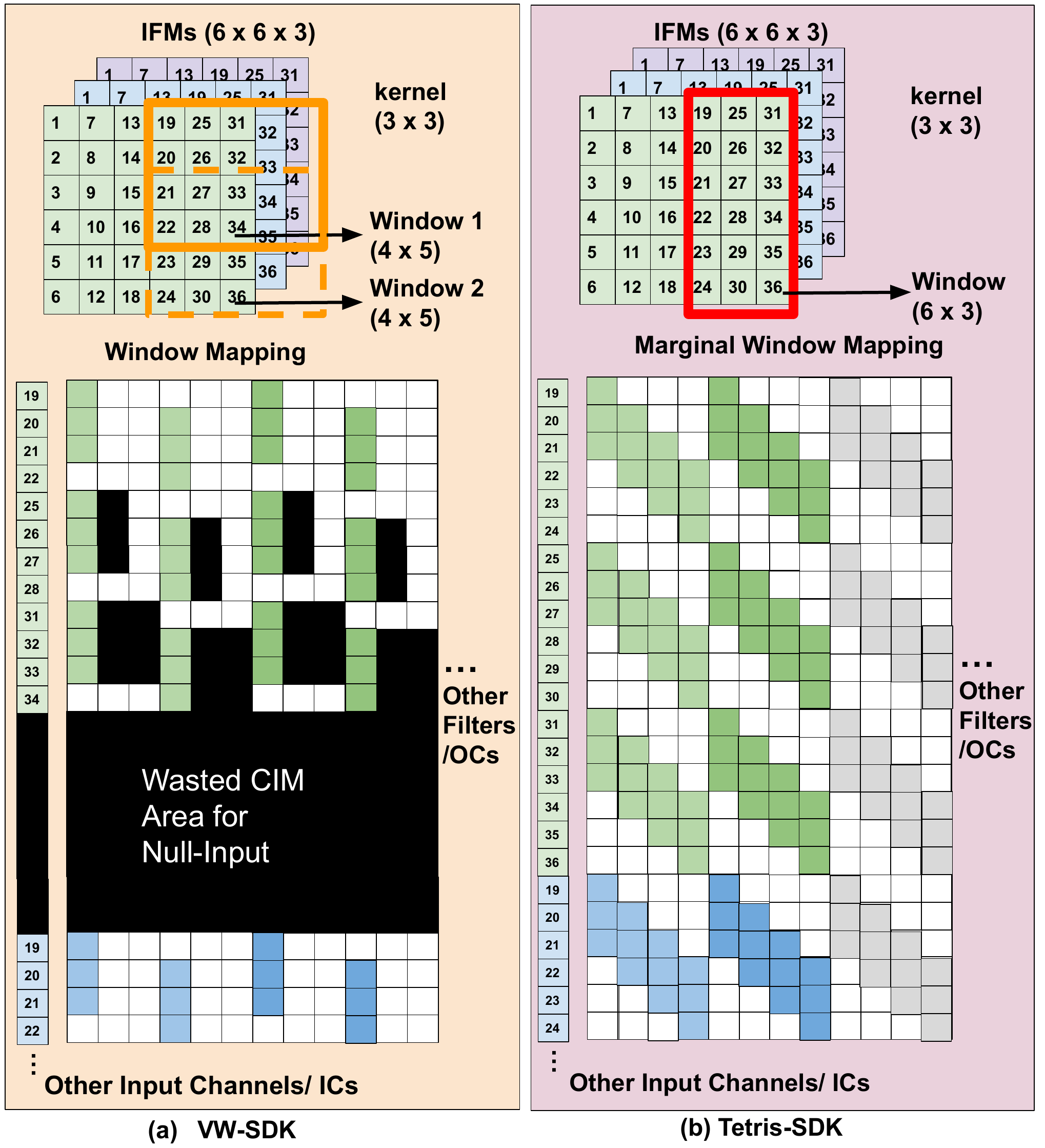}
    \caption{Illustration of marginal space mapping in (a) VW-SDK and (b) Tetris-SDK.
    }
    \label{fig:marginal_window_detail}
\end{figure}

As depicted in Fig. \ref{fig:cnn8-3}(b), two marginal windows are generated to adequately cover the borders of IFMs, instead of four $5\times 6$ parallel windows. Consequently, the computing cycles decrease by $2$.
Fig. \ref{fig:marginal_window_detail} further illustrates the efficacy of the proposed marginal window with CIM array mapping.
Specifically, VW/VWC-SDK needs 2 parallel windows to traverse the borders of the IFMs whereas Tetris-SDK merely utilizes one window. The former incurs wasted area in CIM array due to the null input when the window slides out of the border. 

\begin{algorithm}[hbt!]
\fontsize{7.5}{10}\selectfont
\caption{Finding Marginal Optimized Window}\label{alg:marginal}
\KwData{$I$, $K$, $IC$, $OC$, $AR$, $AC$, $SW$}
\KwResult{$N_{marginal}$, \{Marginal\_Window ($MW$)\}}
$Image\_Marginal_w \gets (I-MW_w)\%(MW_w - K +1)$\;
$Image\_Marginal_h \gets (I-MW_h)\%(MW_h - K +1)$\;
$N_{MWx} \gets 0$\;
$N_{MWy} \gets 0$\;

\If{$Image\_Marginal_w \neq 0$}{
      $MW_w = Image\_Marginal_w + K - 1 $\;
      $MW_h = (SW_w \times SW_h) // MW_w $\;
      $N_{MWx} = \lceil(I/MW_h)\rceil$\;
      }
\If{$Image\_Marginal_h \neq 0$}{
      $MW_h = Image\_Marginal_h + K - 1 $\;
      $MW_w = (SW_w \times SW_h) // MW_h $\;
      $N_{MWy} = \lceil(I/MW_w)\rceil$\;
      }
$N_{marginal} = N_{MWx} + N_{MWy}$

\end{algorithm}

\subsection{A Depth-Optimal Window}
\textcolor{black}{Although VW-SDK accounts for channel partitions, the window size it generates remains identical regardless of channel partition. On the other hand, VWC prunes the residual channels to adapt to channel partitions. However, it does not apply to all network layers. In contrast, Algorithm ~\ref{alg:dep} produces a depth-optimal window that adapts to various channel depths. Following the previous algorithms, firstly we determine the number of remaining channels differing from the prior partition. Subsequently, we adjust the window shapes based on the new channel partitions.}
The procedure repeats with channel pruning to fit all the channels within the CIM macro until all channels are processed. The pruned network will be tested to ensure accuracy loss is minimal.
Prior arts \cite{pruning1}\cite{pruning3} have shown that appropriate channel pruning can remove unnecessary channels with minimal impact on accuracy. 
Our experiment shows the window pruning merely causes a $0.03 \%$ loss in accuracy over MNIST if we prune one channel from the CNN8-3 layer in Fig. \ref{fig:cnn8-3}.

\begin{algorithm}[hbt!]
\fontsize{7.5}{10}\selectfont
\caption{Finding Tetris Window for Variable Depth}\label{alg:dep}
\KwData{$I$, $K$, $IC$, $OC$, $AR$, $AC$, $PW\times IC_t\times OC_t$, $PC$}
\KwResult{Depth\_Window ($DW$), Computing\_Cycles ($CC$), DONE}
$Pruned\_channel \gets 0$\;
$Remaining\_channel \gets PC$\;
$Found\_DO \gets False$\;
$Max\_conv \gets AC/ OC$\;

\While{$(Found\_DO \neq True) \& (Remaining\_channel \neq 0)$}{
  $Remaining\_channel -= 1 $\;
  factorize $Max\_conv$ using square-root\;
  \For {factor $(a, b)$ of $Max\_conv$}
   {\If{$\min(a + K -1)\times(b + K -1)\times Remaining\_channel \leq AR$}{
      $DW_w \gets a$\;
      $DW_h \gets b$\;
      $Found\_DO \gets True$\;
      \Return{$DW, CC$}\;
    }
}$Pruned\_channel \gets Pruned\_channel + 1$\;
}
\If{$Remaining\_channel = 0$}{$DONE \gets True$}

\end{algorithm}

As shown in Fig. \ref{fig:cnn8-3}(b), the remaining channels are 15 in the CNN8-3 layer after applying the square-inclined window and marginal window in the first tile. We can prune one input channel so that 16 windows of $6\times6$ are obtained \textcolor{black}{with respect to 14 input channels}, reducing the overall computing cycles from $48$ to $38$ compared to VW-SDK.

%% file: section/result.tex
\textcolor{black}{
The total number of computing cycles in a CIM-based system is determined by several factors, including the network architecture, the size of the CIM array, and the choice of mapping algorithms. To assess the advantages of \textcolor{black}{Tetris-SDK and }TetrisG-SDK, we perform experiments on a wide range of variables including different network architectures, CIM array sizes, and mapping algorithms.
By systematically varying these parameters, we aim to provide a comprehensive analysis of the impact of each factor on the overall computing cycles and demonstrate the efficacy of \textcolor{black}{Tetris-SDK and }TetrisG-SDK in optimizing the mapping process. 
Furthermore, we open-source our code and make it available at https://github.com/SybilD/TetrisG-SDK}. 
\textcolor{black}{\subsection{Experimental Setup}}
\subsubsection{Network Benchmark}
\textcolor{black}{
In our experiment, we employ CNN8 \cite{vwsdk}, GoogLeNet Inception modules \cite{inception}, and \textcolor{black}{DenseNet40} \cite{densenet40} to assess the performance and efficacy of the proposed mapping algorithm. These network models represent a variety of sizes and application types, providing an appropriate evaluation of mapping algorithms' efficiency. 
}

\begin{figure}[h!]
    \centering
    \includegraphics[width=\linewidth, height=10cm]{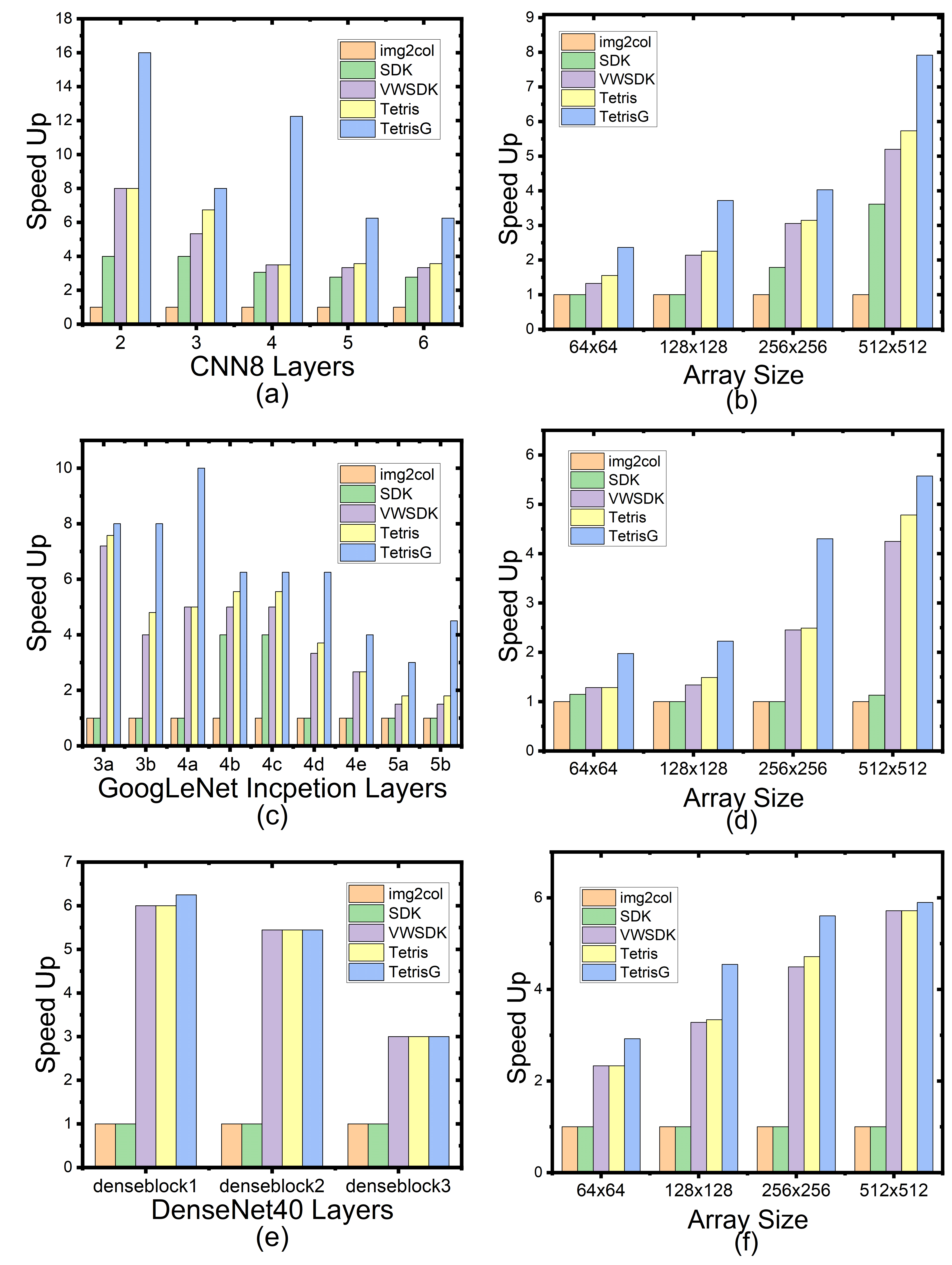}
    \caption{\textcolor{black}{Illustration of speed-up for CNN8, GoogLeNet Inception, and DenseNet40 by Tetris-SDK and TetrisG-SDK across different layers and different array sizes.
    }}
    \label{speedup_all}
\end{figure}

\begin{figure}[h!]
    \centering
    \includegraphics[scale = 0.45]{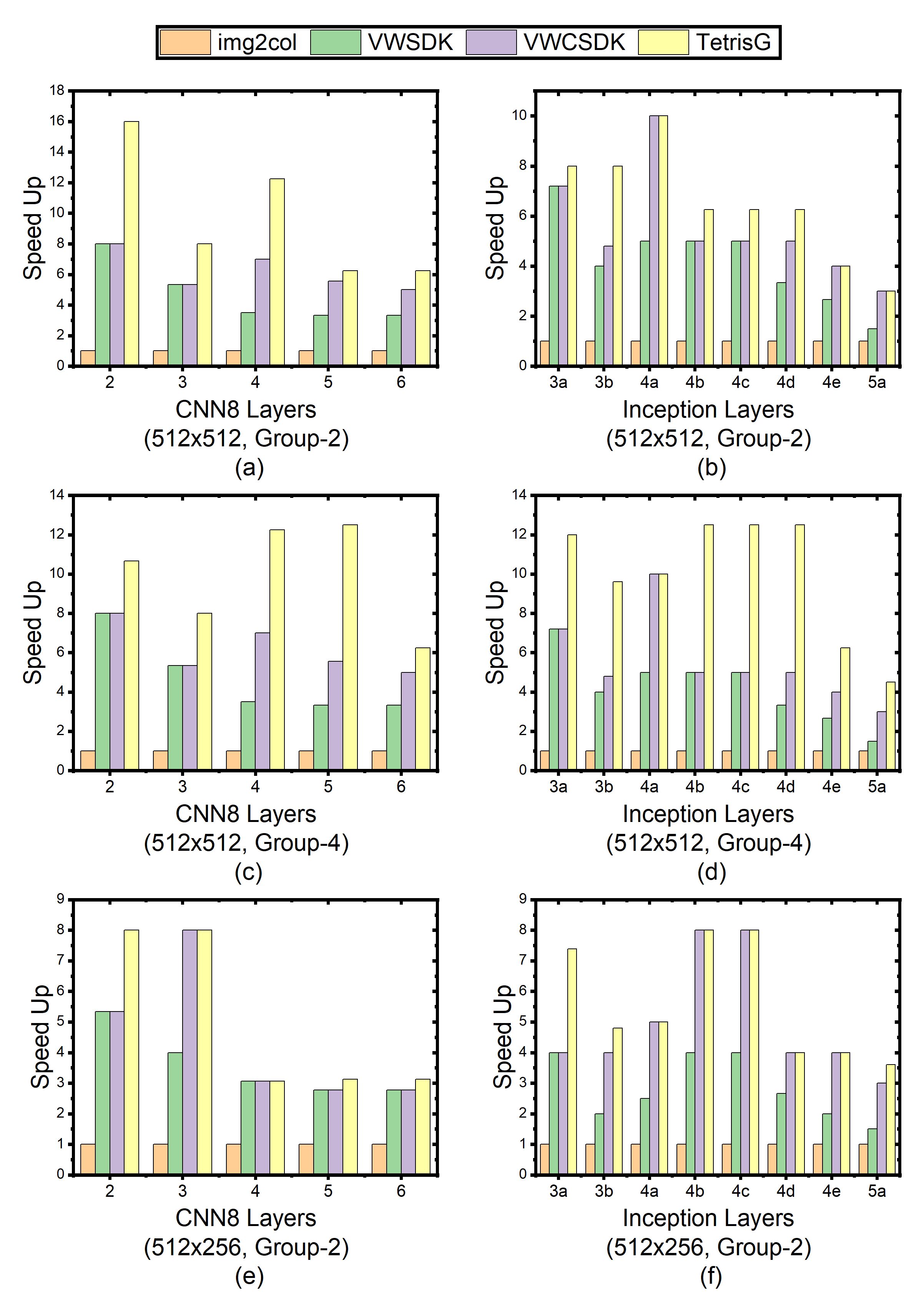}
    \caption{\textcolor{black}{Illustration of speed-up by TetrisG-SDK for benchmarking networks and layers with group convolutions.}}
    \label{fig:multi_speed}
\end{figure}

\subsubsection{Training}

\textcolor{black}{
By adopting the four proposed algorithms, TetrisG-SDK performs a series of computations to determine the minimum computing cycles and the optimal set of window shapes for the given configuration.
In scenarios where the grouped convolutions are utilized, we first train the network using grouped convolutions. This involves modifying the network architecture by introducing the group number parameter in the Conv2D layers in our Pytorch training framework. 
To evaluate the efficacy of grouped convolutions and the impact on accuracy, we perform experiments on benchmarking network architectures, while training with popular image classification datasets such as MNIST, Cifar-10, and Tiny ImageNet. 
}

\subsubsection{Multi-bit Weights}



\textcolor{black}{During the evaluation process, it is assumed that the multi-bit weights are handled via mapping with consecutive columns in CIM arrays as shown in \cite{neurosim}\cite{dong2020}\cite{vwsdk}\cite{vwcsdk}. An alternative implementation is slicing the multi-bit weights into multiple single-bit weights so that each bit can be stored in a specific CIM array \cite{yu2021compute}.} 

\begin{table*}[]
\fontsize{7.8}{10}\selectfont
\centering
\caption{Mapping Results of Various Networks with VW-SDK, VWC-SDK, Tetris-SDK and TetrisG-SDK}
\begin{tabular}{ccccccc}
\hline
\multirow{3}{*}{Layer}                   & IFMs                                                   & Kernel                                            & VW-SDK     & VWC-SDK                                                                    & Tetris-SDK      & TetrisG-SDK                                                                                                                                                                        \\
                                         & \multicolumn{1}{c}{\multirow{2}{*}{($I_w \times I_h$)}} & \multicolumn{1}{c}{($K \times IC$} & \multicolumn{1}{c}{($PW_w \times PW_h$}        & \multicolumn{1}{c}{($PW_w \times PW_h$}               & \multicolumn{1}{c}{($PW_w \times PW_h$}                  & \multicolumn{1}{c}{($PW_w \times PW_h$}                                                                                                                                             \\
                                         & \multicolumn{1}{c}{}                                    & \multicolumn{1}{c}{$\times OC$)}   & \multicolumn{1}{c}{$\times IC_t \times OC_t$)} & \multicolumn{1}{c}{$\times IC_t \times OC_t$)}    & \multicolumn{1}{c}{$\times IC_t \times OC_t$)}     & \multicolumn{1}{c}{$\times IC_t \times OC_t$)}  \\ \hline
\multicolumn{7}{c}{\textbf{CNN8}}                                                                                                     \\ \hline
2                                               & 18x18                                                      & 3x24x32                     & 10x4x12x32                                               &  10x4x12x32                                                              & 6x6x12x32                                                & 6x10x8x16                                                                                        \\
3                                               & 18x18                                                      & 3x32x32                     & 8x4x16x32                                                & 8x4x16x32                                                                                      & 5x6x17x32, 6x6x14x32                                     & 10x6x8x16                                                                                        \\
4                                               & 9x9                                                        & 3x32x64                     & 9x3x18x64                                                & 9x3x18x73                                                                                      & 9x3x18x64                                                & 6x5x16x32, 5x6x16x32, 10x3x16x32                                                                 \\
5                                               & 7x7                                                        & 3x64x64                     & 7x3x24x64                                                & 7x4x18x51                                                                                      & 7x3x24x64, 6x4x16x64                                     & 7x4x18x32, 9x3x18x32, 6x6x14x32                                                                  \\
6                                               & 7x7                                                        & 3x64x64                     & 7x3x24x64                                                &  7x3x24x64                                                             & 7x3x24x64, 6x4x16x64                                     & 7x4x18x32, 9x3x18x32, 6x6x14x32                                                                  \\
7                                               & 5x5                                                        & 5x64x256                    & 5x5x64x256                                               & 5x5x60x256                                                                                     & 5x5x64x256                                               & 5x5x32x128                                                                                       \\ \hline
\multicolumn{3}{c}{total cycles}                                                                                   & 128                                                      & 109                                                                                                  & 116                                                      & 84                                                                                               \\ \hline
\multicolumn{7}{c}{\textbf{GoogLeNet Inception}}                                                                                                \\ \hline
3a                                              & 28x28                                                      & 5x16x32                     & 9x7x8x32                                                 & 9x7x8x34                                                                                       & 9x7x8x32, 8x8x8x32                                              & 8x8x8x16                                                                                             \\
3b                                              & 28x28                                                      & 5x32x96                     & 9x5x11x96                                                & 9x5x11x102                                                                                     & 9x5x11x96, 8x5x12x96                                            & 9x6x9x48, 8x7x9x48                                                                                   \\
4a                                              & 14x14                                                      & 5x16x48                     & 9x6x9x48                                                 & 9x6x9x51                                                                                       & 9x6x9x48                                                        & 9x7x8x24, 12x5x8x24                                                                                  \\
4b                                              & 14x14                                                      & 5x24x64                     & 7x6x12x64                                                & 7x6x12x85                                                                                      & 7x6x12x64, 5x8x12x64                                            & 9x7x8x32, 12x5x8x32, 8x8x4x32                                                                        \\
4c                                              & 14x14                                                      & 5x24x64                     & 7x6x12x64                                                & 7x6x12x85                                                             & 7x6x12x64, 5x8x12x64                                            & 9x7x8x32, 12x5x8x32, 8x8x4x32                                                                        \\
4d                                              & 14x14                                                      & 5x32x64                     & 9x5x11x64                                                & 9x5x11x102                                                                                     & 9x5x11x64, 6x8x10x64, 7x6x10x64                                 & 9x7x8x32, 12x5x8x32                                                                                  \\
4e                                              & 14x14                                                      & 5x32x128                    & 6x6x14x128                                               & 6x6x14x128                                                                                     & 6x6x14x128                                                      & 8x6x10x64, 6x8x10x64                                                                                 \\
5a                                              & 7x7                                                        & 5x32x128                    & 6x5x17x128                                               & 7x5x14x170                                                                                     & 6x5x17x128, 5x6x17x128, 6x6x14x128                              & 6x5x16x64, 5x6x16x64                                                                                 \\ \hline
\multicolumn{3}{c}{total cycles}                                                                                   & 627                                                      & {\color[HTML]{212121} 506}                                                                           & 557                                                             & 470                                                                                                  \\ \hline
\end{tabular}
\label{Table: windows}
\end{table*}

\begin{table*}[]
\fontsize{6.8}{10}\selectfont
\centering
\caption{\textcolor{black}{Comparison on Accuracy and Computing Cycles of Conventional Convolutions and Grouped Convolutions}}
\begin{tabular}{cccccccccccccc}
\hline
\multirow{4}{*}{Networks} & \multirow{4}{*}{Dataset} & \multicolumn{2}{c}{\multirow{3}{*}{VW-SDK}} & \multicolumn{2}{c}{\multirow{3}{*}{VWC-SDK}} & \multicolumn{4}{c}{\multirow{2}{*}{Tetris-SDK}}              & \multicolumn{4}{c}{\multirow{2}{*}{TetrisG-SDK}}             \\
                          &                          & \multicolumn{2}{c}{}                            & \multicolumn{2}{c}{}                             & \multicolumn{4}{c}{}                                     & \multicolumn{4}{c}{}                                     \\
                          &                          & \multicolumn{2}{c}{}                            & \multicolumn{2}{c}{}                             & \multicolumn{2}{c}{VW-SDK Training} & \multicolumn{2}{c}{VWC-SDK Training} & \multicolumn{2}{c}{VW-SDK Training} & \multicolumn{2}{c}{VWC-SDK Training} \\
                          &                          & Accuracy                 & CCs                  & Accuracy                  & CCs                  & Accuracy       & CCs       & Accuracy       & CCs        & Accuracy       & CCs       & Accuracy        & CCs       \\ \hline
CNN8                      & SVHN                     & 94.21\%                  & 168                  & 94.66\%                   & 135                  & 94.21\%        & 158       & 94.66\%        & 126        & 94.78\%        & 126       & 94.46\%         & 114       \\ \hline
DenseNet40                  & CIFAR-10                  & 81.67\%                  & 1488                 & 81.67\%                   & 1488                 & 81.67\%        & 1416      & 81.67\%        & 1416       & 83.68\%        & 1164      & 83.68\%         & 1164      \\ \hline
Inception                 & Tiny ImageNet            & 44.11\%                  & 1059                 & 44.41\%                   & 614                  & 44.11\%        & 1035      & 44.41\%        & 612        & 44.12\%        & 718       & 44.89\%         & 476       \\ \hline

\end{tabular}
\label{Table:accuracy}
\end{table*}

\textcolor{black}{\subsection{Comparison of Tetris-SDK and VW-SDK}}


\textcolor{black}{
We evaluate our proposed algorithm with a variety of benchmark neural networks including CNN8, GoogLeNet Inception, and DenseNet40. 
}
The mapping results for CNN8 and GoogLeNet Inception Network using $512 \times 512$ CIM array are displayed in Table \ref{Table: windows}. Note that the first layer of CNN8 is excluded as most prior arts \cite{quant1}\cite{quant2} do not quantize and accelerate it. 
\textcolor{black}{From Table \ref{Table: windows}, we can see that Tetris-SDK results in less cycle count compared to VW-SDK while TetrisG-SDK generates the least. 
}

Fig. \ref{speedup_all}(a), (c), and (e) show the speed up of Tetris-SDK compared to img2col, SDK and VW-SDK over a variety of networks and CNN layers.
We can see that SDK fails to improve most of the Inception layers as the available kernel size is constrained by the CIM array row numbers. VW-SDK is able to improve SDK by channel partition but the computing cycles are not optimal due to the issues elaborated in Sections I and II. Tetris-SDK reduces the computing cycles for most of the networks, except for those cases where the number of input channels is too small to be optimized. 

\textcolor{black}{Fig. \ref{speedup_all}(b), (d), and (f) demonstrate that Tetris-SDK is effective in improving computing cycles across a range of array sizes from 64 × 64 to 512 × 512. In contrast, SDK fails to improve most layers due to its approach of accommodating entire channels. Specifically, for small CIM array sizes and large kernel sizes (e.g., in GoogLeNet Inception model), the rigid shapes of SDK are less effective compared to other mapping methods.}
In general, Tetris-SDK gains higher speed-up with a larger CIM array than the conventional algorithms.
This is caused by the fact that a larger CIM array tends to accommodate more convolutions so that Tetris-SDK can find larger optimal windows to minimize the overall computing cycles.






\textcolor{black}{\subsection{Comparison of TetrisG-SDK and VWC-SDK}}

\subsubsection{Accuracy}

\textcolor{black}{As Table \ref{Table:accuracy} shows, we select a variety of network architectures and datasets to evaluate the efficacy of grouped convolutions. 
\textcolor{black}{For fair comparisons, Tetris-SDK and TetrisG-SDK (not including group convolutions) are trained using the same methodology as VW-SDK and VWC-SDK. Using the VWC-SDK training method, the network topology is derived through the VWC-SDK training framework with channel pruning, while using the VW-SDK training approach, the topology is obtained through the framework without channel pruning.}
The experimental results reveal that TetrisG-SDK generates the least computing cycles. 
Particularly, grouped convolutions exhibit minimal accuracy loss (i.e., at most 0.2\%) compared to VW-SDK, VWC-SDK and Tetris-SDK. It even leads to an accuracy improvement of 2.01\% for DenseNet40 on the CIFAR-10 dataset. \textcolor{black}{This aligns with prior observations that grouping can improve generalization by reducing filter co-adaptation and feature redundancy \cite{dgc}.}
This experiment shows the applicability of grouped convolutions across various workloads. In addition, grouped convolutions can be opted out for use if they downgrade the accuracy by a certain threshold which the users can determine. 
}
\subsubsection{Utilization}
\textcolor{black}{We compare the utilization rate of CIM arrays at marginal space for TetrisG-SDK and VWC-SDK.}
\textcolor{black}{Compared to VWC-SDK, TetrisG-SDK achieves improvements of 1.6×, 2.0×, and 1.5× in mapping at marginal space for CNN8, GoogLeNet Inception, and DenseNet40, respectively.
These results verify the efficiency of the marginal window approach implemented in TetrisG-SDK, which optimizes the utilization of available CIM resources. 
}

\subsubsection{Computing Cycles}


In Fig. \ref{fig:multi_speed}, we compare the speed-up achieved by TetrisG-SDK against conventional methods across various networks and CNN layers. We prune the channels of CNN8 by 2.8\% while keeping Inception as it is. TetrisG-SDK demonstrates improvements in almost all layers, including cases where VW-SDK fails to reduce the computing cycles. This can be attributed to the utilization of grouped convolutions in TetrisG-SDK, which effectively reduces the computational workload. 
\textcolor{black}{Generally, with multiple grouping, TetrisG-SDK achieves significant speed-up. \textcolor{black}{For $512\times512$ CIM array}, TetrisG-SDK yields an average speed-up of $1.5\times$ for CNN8, $1.3\times$ for GoogLeNet Inception, and $2\times$ for DenseNet40 against VW-SDK, respectively.}


\textcolor{black}{Fig. \ref{speedup_all}(b), (d), and (f) have demonstrated the mapping performance of TetrisG-SDK with different CIM array sizes. This improvement is further enhanced by the reduction of computation with grouping convolutions.}
\textcolor{black}{Besides, compared to VWC-SDK on CNN8 with a $256 \times 512$ CIM, TetrisG-SDK improves the computing cycles by $1.7 \times$ with 2-grouped convolutions when the network undergoes one round of residual-channel-pruning and $1.5 \times$ when it goes through two rounds. This demonstrates the efficacy of TetrisG-SDK over the state-of-the-art mapping methodology. 
\textcolor{black}{Note that the performance of VWC-SDK is based on the reported numbers in \cite{vwcsdk}.}
}
\textcolor{black}{We further map MobileNet \cite{mobilenet}, a typical lightweight CNN, using our framework. TetrisG achieves the lowest total cycle count on MobileNet, delivering a $2.8\times$ speedup compared to img2col. In particular, TetrisG demonstrates the same performance as VWC-SDK, as the mixture of depthwise and pointwise layers limits cross-channel reuse and parallelism, thereby constraining opportunities for further enhancement.}

\subsection{DNN+NeuroSim Simulation}
\begin{figure}[h!]
    \centering
    \includegraphics[width=\linewidth, height=7cm]{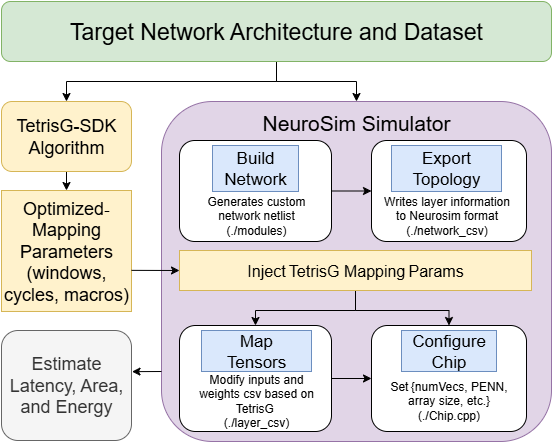}
    \caption{\textcolor{black}{Deployment of TetrisG-SDK with DNN+NeuroSim.
    }}
    \label{neurosim-flowchart}
\end{figure}

\textcolor{black}{We integrate DNN+NeuroSim\cite{neurosim}, a well-known, open-source, and reconfigurable simulator to validate the efficacy of the proposed TetrisG-SDK mapping method from application and system perspectives. Accordingly, we implement CNN8, DenseNet40, and GoogLeNet Inception on the simulator. The workflow is depicted in Fig. \ref{neurosim-flowchart}. Specifically, we select the desired network architecture and dataset for both TetrisG-SDK and NeuroSim. TetrisG-SDK analyzes the neural network, outputs optimized window shapes for mapping, and calculates computing cycles, which are subsequently supplied into DNN+NeuroSim. In NeuroSim, adjustments are made to the files under ./modules and ./network\_csv based on the model and dataset for training. With the results of TetrisG-SDK, we then implement the mapping by updating the layer\_csv files to include inputs and weights. It is important to note that the input and weight data need to be prepared offline to adapt to the mapping strategy. Besides, we modify DNN+NeuroSim parameters and functions to set the number of input vectors and tune AR/AC cycles. After completing these configurations, we execute the inference, producing the latency and energy metrics. For the simulation, we select the cell type as SRAM, the array size as $512\times512$, and the operation mode as parallel. Throughout the implementation, we maintain a clock frequency of 1 GHz and an operating temperature of 300K, based on a technology node of 22 nm CMOS technology.
}

\begin{figure}[h!]
    \centering
    \includegraphics[width=\linewidth, height=10cm]{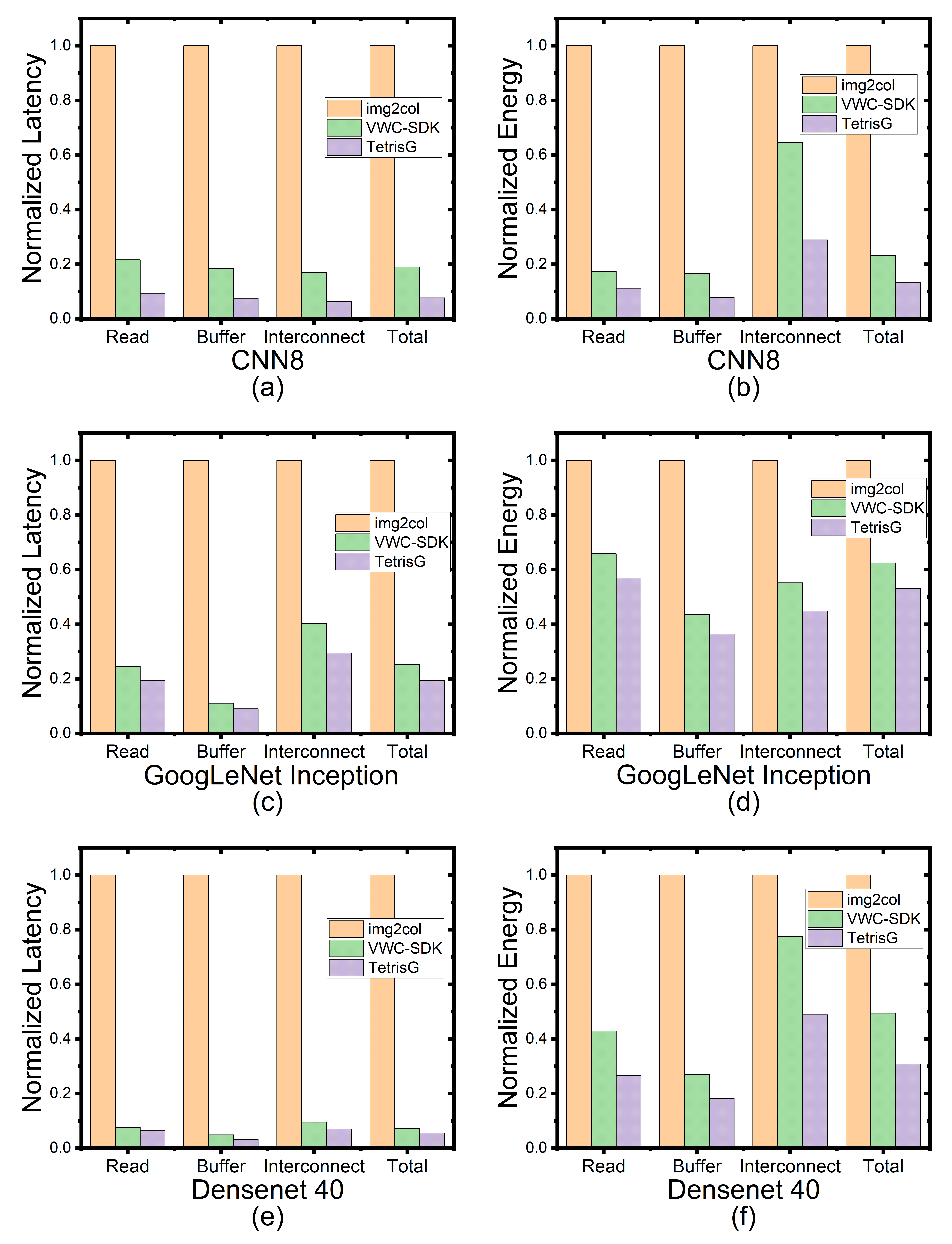}
    \caption{\textcolor{black}{Normalized latency and dynamic energy consumption of CNN8, GoogLeNet Inception, and DenseNet40 on the DNN+NeuroSim simulator.}
    }
    \label{neurosim-result-all}
\end{figure}

\begin{figure}[t]
\centering
\includegraphics[width=\linewidth]{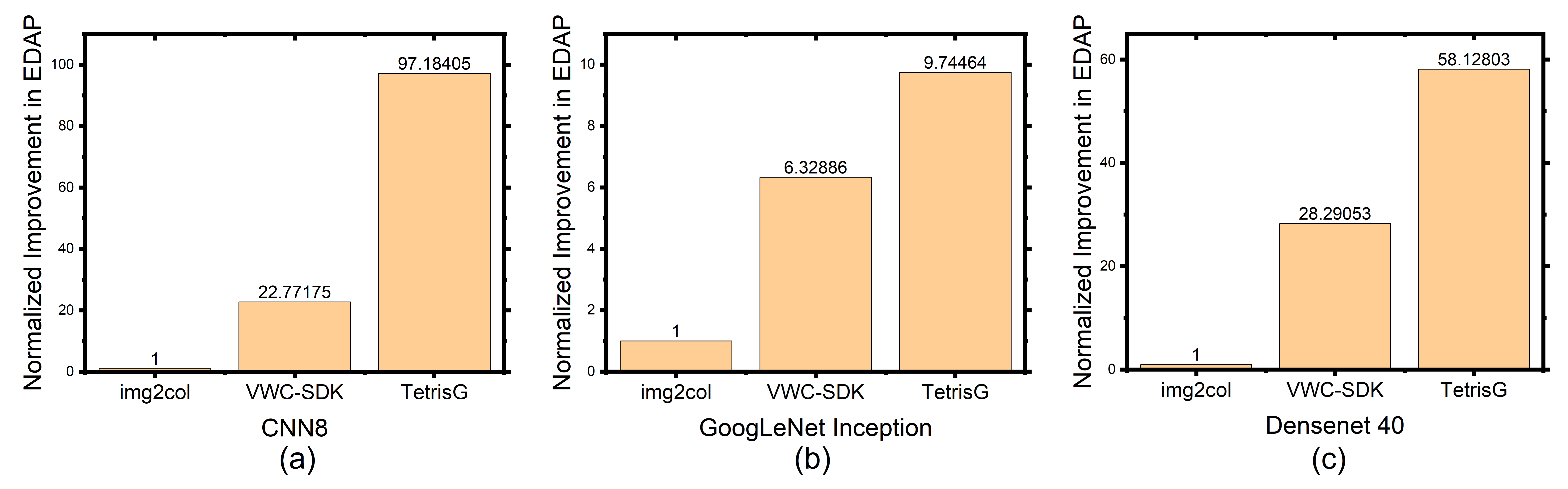}
\caption{\textcolor{black}{Normalized EDAP of img2col, VWC-SDK and TetrisG‑SDK on (a) CNN8, (b) Inception, and (c) DenseNet40.}}
\label{fig:singlemacro-edap}
\end{figure}

\begin{figure}[t]
\centering
\includegraphics[width=\linewidth]{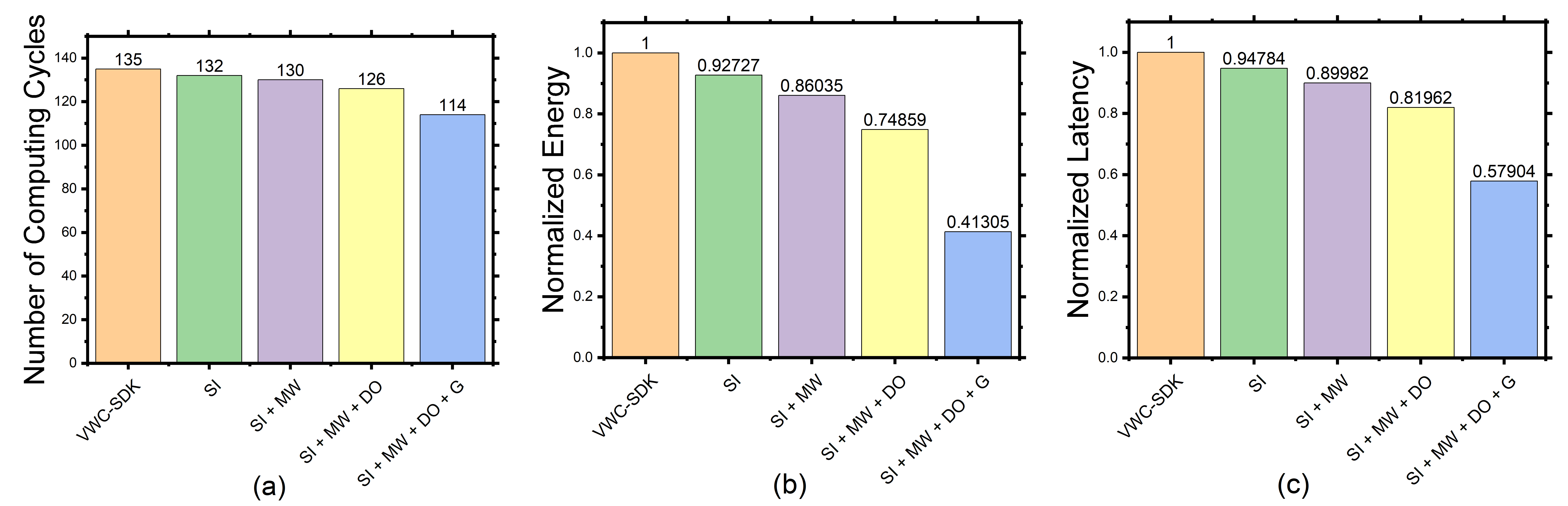}
\caption{\textcolor{black}{Illustration of stepwise ablation study on CNN8 mapping: (a) total computing cycles, (b) normalized energy, and (c) normalized latency.}}
\label{fig:ablation_cnn8}
\end{figure}

\begin{figure}[t]
\centering
\includegraphics[width=\linewidth]{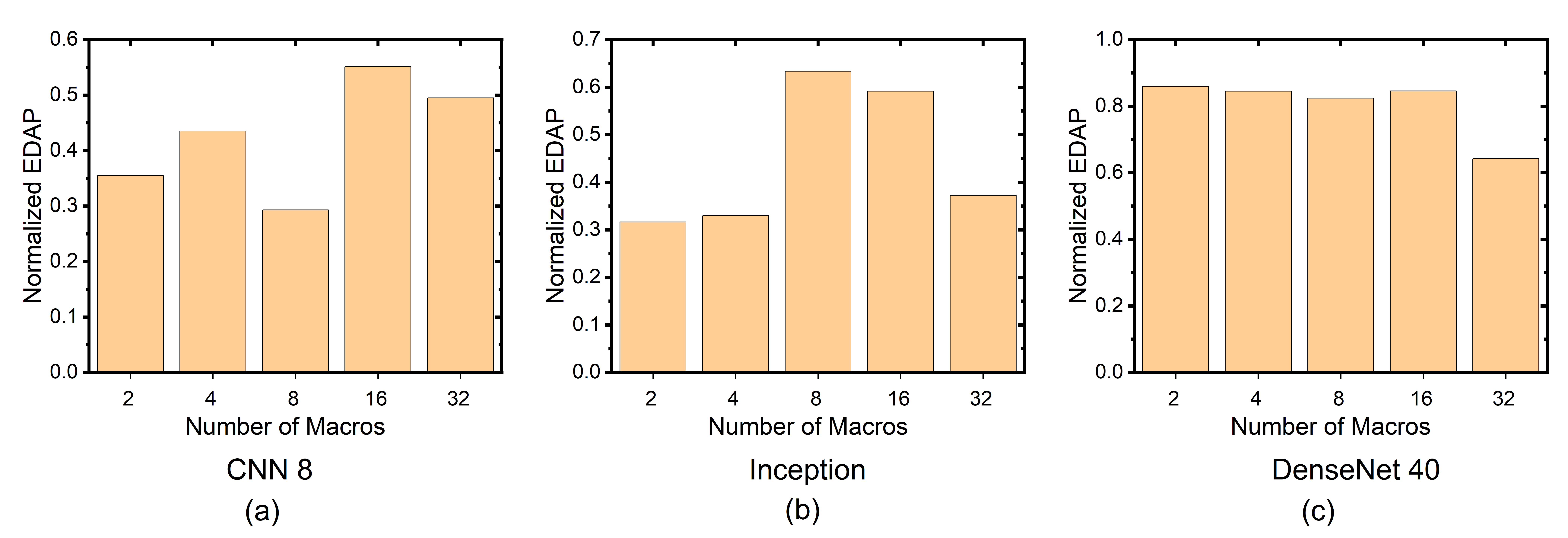}
\caption{\textcolor{black}{Normalized EDAP of TetrisG‑SDK across multiple macros on (a) CNN8, (b) Inception, and (c) DenseNet40.}}
\label{fig:macro-edap}
\end{figure}

\textcolor{black}{
As depicted in Fig. \ref{neurosim-result-all}, compared to VWC-SDK, TetrisG-SDK reduces the latency and energy by $2.4\times$ and $1.7\times$ for CNN8, $1.3\times$ and $1.2\times$ for GoogLeNet Inception, and $1.3\times$ and $1.6\times$ for DenseNet40, respectively. 
\textcolor{black}{Moreover, it achieves EDAP improvements of $97.2\times$, $9.7\times$, $58.1\times$ over img2col and $4.27\times$, $1.54\times$, $2.06\times$ over VWC-SDK for CNN8, Inception, and DenseNet40, respectively (Fig. \ref{fig:singlemacro-edap}). }
This analysis includes the impact of various peripheral components, such as overhead from reading, buffering, and interconnection. The decreased computing cycles and optimized window shapes for mapping lead to fewer read operations and thus lower energy consumption from ADCs and the accumulation units. 
}

We further perform a stepwise ablation study on CNN8 under the same configuration. Beginning with the VWC-SDK baseline, we incrementally enable square-inclined window selection (SI), marginal window search (MW), depth-optimal handling (DO), and grouped convolution (G). As shown in Fig.~\ref{fig:ablation_cnn8}, grouped convolution provides the most significant performance gain by reducing per-group channel dimensions and alleviating AR/AC tiling constraints, thereby enlarging the parallel window size.

\subsection{Effect of Macro Parallelism Across Networks}

Fig.~\ref{fig:macro-edap} visualizes the EDAP of TetrisG‑SDK normalized to Tetris‑SDK under the same macro budget $P$. Each macro is a $64{\times}64$ synaptic array modeled in DNN+NeuroSim. For every $P$, the macro‑grid search (Alg.~\ref{alg:macro}) evaluates all feasible $(r,c)$ layouts and selects the one that minimizes $CC_{\text{multi}}$ before we evaluate system‑level EDAP. Across all networks, TetrisG-SDK consistently reduces computing cycles and hence lowers EDAP under equal hardware constraints.
For CNN8, the optimal EDAP occurs with 8 macros and yields a 70\% reduction. This improvement arises because the search activates only 6 macros to achieve the minimum $CC_{\text{multi}}$ and power-gates the remaining 2. In contrast, Tetris-SDK requires all 8 macros to be active.
The optimal EDAP for Inception is along with 2 macros, delivering a 68\% reduction. \textcolor{black}{Performance gain is slightly smaller with 8 and 16 macros due to the Inception-3b layer, which dominates the mapping performance. Specifically, TetrisG-SDK obtains the same cycle count as the non‑grouped baseline but uses fewer active macros. Thereby, EDAP benefits merely from the reduced number of active macros.}
\textcolor{black}{For DenseNet40, EDAP improves consistently as the macro budget grows and reaches its minimum along with 32 macros, leading to a 36\% reduction. The dominant bottleneck is the input‑channel partition $IC_t$. Grouped convolutions alleviate this by shrinking the per‑group channel depths $(IC/G,\,OC/G)$, which enables the search to distribute groups across the macro grid and enlarge the windows so that $N^{(P)}_{\text{windows}}$ in Eq. (6) drops.} \textcolor{black}{The substantial improvement with 32 macros arises because TetrisG‑SDK reduces computing cycles and the required number of active macros. This trend persists for larger budgets as long as (i) the grid search generates configurations with idle macros, and (ii) the mapping windows keep the cycle counts low.}
\textcolor{black}{In summary, leveraging macro parallelism with grouped convolutions yields substantial EDAP improvement given the same hardware budgets.}

%% file: section/conclusion.tex
\textcolor{black}{We presented TetrisG‑SDK, an efficient mapping framework to accelerate convolutions in CIM accelerators. It introduces a macro‑grid search to optimize parallel macro configurations, integrates adaptive windows, and leverages grouped convolutions with negligible accuracy loss.
Implemented in DNN+NeuroSim, TetrisG‑SDK achieves significant reductions in latency and energy of 2.4× and 1.7× for CNN8, 1.3× and 1.2× for Inception, and 1.3× and 1.6× for DenseNet40.
By exploiting macro parallelism, our approach also reduces the EDAP by 70\%, 68\%, and 36\% across these networks, respectively. 
These results reveal that TetrisG-SDK is a highly effective strategy for mapping and accelerating CNN workloads on CIM-based architectures.
}